\documentclass[prd, reprint, amsmath, amssymb, superscriptaddress, nofootinbib, showpacs, aps, floatfix]{revtex4-1}

\usepackage[justification=justified,font=small,labelfont=bf]{caption}

\newcommand{\Chi}{\mathrm{X}}

\usepackage{textcase}
\usepackage{graphicx}%
\usepackage{multirow}
\usepackage{float}
\usepackage{epstopdf}
\usepackage[utf8]{inputenc}
\usepackage[english]{babel}
\usepackage{footnote}
\usepackage{xcolor}
\usepackage{textcomp}
\usepackage{color} 
\usepackage{orcidlink}
\usepackage{soul}
\usepackage{siunitx}
\usepackage{comment}
\usepackage{subcaption}

\usepackage[capitalise,noabbrev]{cleveref}

\usepackage[skip=6pt plus1pt, indent=12pt]{parskip}
\usepackage{silence}
\begin{document}

\title{Simulation of Muon-induced Backgrounds for the Colorado Underground Research Institute (CURIE)}%

\author{Dakota K. Keblbeck \orcidlink{0000-0002-4087-9326}}\thanks{Corresponding author}
\affiliation{Department of Physics, Colorado School of Mines, Golden, Colorado 80401, USA}

\author{Eric Mayotte \orcidlink{0000-0003-2618-9166}}
\affiliation{Department of Physics, Colorado School of Mines, Golden, Colorado 80401, USA}

\author{Uwe Greife \orcidlink{0000-0003-2906-8466}}
\affiliation{Department of Physics, Colorado School of Mines, Golden, Colorado 80401, USA}

\author{Kyle G. Leach \orcidlink{0000-0002-4751-1698}}
\affiliation{Department of Physics, Colorado School of Mines, Golden, Colorado 80401, USA}

\author{Wouter Van De Pontseele \orcidlink{0000-0003-3422-313}}
\affiliation{Department of Physics, Colorado School of Mines, Golden, Colorado 80401, USA}

\author{Caitlyn Stone-Whitehead \orcidlink{0009-0000-5396-7328}}
\affiliation{Department of Physics, Colorado School of Mines, Golden, Colorado 80401, USA}

\author{Luke Wanner \orcidlink{0009-0006-2221-1740}}
\affiliation{Department of Physics, Colorado School of Mines, Golden, Colorado 80401, USA}

\author{Grace Wagner \orcidlink{0000-0003-0295-875X}}
\affiliation{Department of Physics, Colorado School of Mines, Golden, Colorado 80401, USA}

\date{\today}%

\begin{abstract}
We present a comprehensive Monte Carlo simulation of muon-induced backgrounds for the Colorado Underground Research Institute (CURIE), a shallow-underground facility with $\approx 415$~m.w.e. overburden. Using coupled \textsc{mute} and \textsc{geant4} frameworks, we characterize the production and transport of muon-induced secondaries through site-specific rock compositions and geometries, establishing a proof-of-concept for high-precision, end-to-end simulations. Our simulations employ angular-dependent muon energy distributions, which improve secondary flux accuracy. For the Subatomic Particle Hideout and Cryolab I research spaces, we predict total muon-induced neutron fluxes of $(8.52 \pm 1.30_{\text{sys}}) \times 10^{-3}$~m$^{-2}$s$^{-1}$ and $(8.86 \pm 1.62_{\text{sys}}) \times 10^{-3}$~m$^{-2}$s$^{-1}$, respectively. Electromagnetic backgrounds are expected to dominate the total flux, with $\gamma$-ray components of $(5.54 \pm 0.70_{\text{sys}}) \times 10^{-1}$~m$^{-2}$s$^{-1}$ and $(6.51 \pm 1.06_{\text{sys}}) \times 10^{-1}$~m$^{-2}$s$^{-1}$ for the Subatomic Particle Hideout and Cryolab I facilities, respectively. Additionally, we develop a Depth-Intensity Relation (DIR) to predict the muon-induced neutron flux as a function of facility depth, which is consistent with measurements across a broad range of underground depths. These results provide quantitative background predictions for experimental design and sensitivity projections at shallow- and deep-underground facilities. They further demonstrate that local geology and overburden geometry influence muon-induced secondary yields and energy spectra, emphasizing the need for site-specific simulations for accurate underground background characterization. Therefore, the simulation framework has been made publicly available at \href{https://doi.org/10.5281/zenodo.17196581}{https://doi.org/10.5281/zenodo.17196581}, for the broader low-background physics community to enable meaningful inter-facility comparisons.
\end{abstract}

\maketitle

\section{Introduction}
Low-background experiments require detailed knowledge of the expected sources of background radiation in order to apply appropriate suppression and rejection techniques, as well as select shielding materials in their design. External sources of experimental backgrounds, mainly cosmic radiation and natural radioactivity, are often dominant contributors to background rates for low-background experiments \cite{formaggio_backgrounds}. One of the simplest methods to suppress the cosmic backgrounds, those created in the atmosphere from particle cascades induced by primary cosmic rays \cite{Kampert2012EASReview}, is to deploy experiments in underground research facilities. These facilities, characterized here as shallow- (shielding $<$ 1 kilometer-water-equivalent (km.w.e.)) and deep-underground (shielding $>$ 1 km.w.e.) provide near-elimination of the surface level cosmic backgrounds due to the overhead rock. For underground laboratories with depths greater than a few 10s of meters, the only non-negligible, surviving cosmic radiation is the cosmogenic muon flux \cite{workman}. Depending on the depth of the underground laboratory, the attenuation of the muon background relative to sea level varies by up to seven orders of magnitude \cite{Guo_2021}.

The underground muon background at a given location is generally irreducible and must be taken into account either through background subtraction techniques or by using in situ rejection techniques, such as coincidence or anticoincidence tagging \cite{Freund2016GerdaMuonVeto, THOMAS201347}. If the local muon flux and their angular and energy distributions are well characterized, they provide the basis for designing shielding to suppress muon-induced backgrounds and for implementing effective muon veto systems for a given detector setup. Additionally, as surface-level muons propagate to depth, they induce electromagnetic and hadronic particle showers along their track~\cite{workman}. Although these muon-induced secondaries are largely absorbed, those produced in the surrounding rock near the underground lab spaces can cross from the rock into the cavern, adding to background radiation.

One of the main issues for low-background experiments, particularly those that require extremely low background event rates, such as neutrinoless double beta-decay ($0\nu\beta\beta$) or dark matter (DM) searches, is the neutron background \cite{amore_neutron, dark_matter_neutron}. Neutrons are more penetrating than $\gamma$-rays and charged particles and interact differently within common $\gamma$-ray shielding materials, and thus require neutron-specific suppression methods, like shielding materials with high hydrogen content (e.g. polyethylene) or enhanced neutron capture (e.g. gadolinium). The total underground neutron background arises from ($\alpha, n$) reactions induced by the $\alpha$-particles from the natural radioactive background in rock, cosmogenic muon interactions in the surrounding rock and materials, and from spontaneous fission from $^{238}$U and $^{232}$Th~\cite{formaggio_backgrounds, KUDRYAVTSEV2020164095}. 

The background signals of interest, from neutron and neutron-induced events, vary depending on the experiment and detector. For example, $\gamma$-rays emitted from inelastic scattering of neutrons ($n, n'\gamma$) can deposit energies in the region of interest for $0\nu\beta\beta$ searches~\cite{neutron_inelastic}. Furthermore, neutron activation of isotopes, such as the production of $^{137}$Xe from $^{136}$Xe, can create a background in the region of interest for $0\nu\beta\beta$ experiments, like nEXO and NEXT \cite{Albert2018, Rogers_2020}.  For dark matter experiments, like those searching for Weakly Interacting Massive Particles (WIMPs), neutron elastic scattering can produce signals which mimic WIMP elastic scattering off a nucleus \cite{Aprile_2013, MEI2009651}.

The neutron flux from the ($\alpha,n$) reactions depends on the alpha activity of the natural radioactive content in the rock and the amount of light target nuclei within the laboratory and experimental setup (the ($\alpha,n$) cross-section for heavy nuclei is suppressed by the Coulomb barrier). The $\alpha$ energies from the radioisotopes are typically $<$ \SI{8}{MeV}, however, the ($\alpha,n$) neutrons can have energies of $>$ \SI{10}{MeV} depending on the material \cite{MEI2009651}. Neutrons from muon-induced events are created in several ways, either directly or indirectly caused by the cosmogenic muons~\cite{workman}. Direct muon-induced neutrons can be generated via muon spallation or negative muon capture. Indirect, muon-induced neutron production can be caused by photon or hadron spallation (or disintegration), where the photons and hadrons are created in muon-induced cascades in the surrounding rock and materials within the experimental setup. Neutrons generated from muon-induced events can be created with energies on the order of \SI{}{GeV} \cite{KUDRYAVTSEV2003688} and are therefore much more penetrating than those created from the natural radioactivity induced ($\alpha,n$) events. For this reason, characterizing the muon-induced neutron background becomes increasingly important for rare-event searches in order to achieve acceptable background levels.

While many DM and $0\nu\beta\beta$ experiments require deep-underground facilities, in part due to their extremely low cosmogenic muon backgrounds, and therefore highly suppressed muon-induced neutron background, many research and development (R\&D) efforts can tolerate moderate muon fluxes and thus require only modest shielding. For instance, radioassay of material purity, for low-background detectors, can be performed in shallow-underground facilities \cite{pnnl}. While a handful of these shallow-underground facilities exist worldwide, only a few are located in the United States. Two of these U.S. facilities are hosted at Department of Energy National Laboratories: Pacific Northwest National Laboratory (PNNL), with an overburden of \SI{30} m.w.e.~\cite{pnnl}, and Fermilab, with an overburden of roughly \SI{225} - \SI{300} m.w.e.~\cite{Figueroa2024, temples}. Both laboratories support active basic and applied science programs. The third facility, the Colorado Underground Research Institute (CURIE), is a new underground lab being developed by the Colorado School of Mines (CSM) \cite{curie}. Contrary to PNNL and Fermilab, CURIE offers easy, horizontal site access and better surface-level muon attenuation, providing a factor of 600-700 reduction~\cite{dakota}. Given the increasing interest and need for these research spaces, along with the increasingly stringent background radiation requirements for low-background detectors, having a well-characterized facility is essential in order to accurately predict and optimize underground experiments and R\&D efforts.

This work presents a quantitative, simulation-based analysis and characterization of the muon-induced secondary backgrounds for the Colorado Underground Research Institute, using \textsc{mute}~\cite{mute, mute2} and \textsc{geant4}~\cite{g4} simulation packages.

\section{Colorado Underground Research Institute}\label{sec:CURIE}
The Colorado Underground Research Institute (CURIE) is a new shallow–underground laboratory developed by the Colorado School of Mines, located within the Edgar Experimental Mine near Idaho Springs, Colorado \cite{curie}. The Edgar Mine has historically served as a teaching and research facility for mining engineering, and now provides the infrastructure for a physics-focused underground laboratory. At CURIE, two primary laboratory spaces are presently in use: the Subatomic Particle Hideout (SPH) and Cryolab I (CLI). The SPH, brought online in 2024, has interior dimensions of roughly \SI{5.2}{m} $\times$ \SI{3.0}{m} $\times$ \SI{3.0}{m}, while the larger CLI space, still under development, measures approximately \SI{12.2}{m} $\times$ \SI{4.9}{m} $\times$ \SI{4.6}{m}.

The facility is classified as a shallow site, with an effective rock overburden corresponding to about \SI{415}{} m.w.e.\ \cite{dakota}. This overburden reduces the muon flux by a factor of $\approx$ 700, relative to sea-level, leaving a residual muon intensity of order $10^{-1}$~m$^{-2}$s$^{-1}$ at laboratory depth. As a result, CURIE provides significant suppression of cosmic-ray muons compared to surface laboratories, while maintaining convenient horizontal access and logistical advantages for research and development activities. These features make CURIE complementary to deeper U.S. facilities, bridging the gap between surface-level facilities and deep-underground experiments requiring greater than 1 km.w.e.\ shielding.

\section{Computational Framework}\label{sec:CompFrame}

The simulation of underground muon-induced backgrounds presents a significant computational challenge due to the need to model both the complex transport of primary muons through realistic overburden and the subsequent production of secondary particles in the laboratory environment. To address this challenge efficiently, we employ a two-stage simulation approach that couples \textsc{mute} v2.0.1~\cite{mute, mute2} and \textsc{geant4}~\cite{g4} v11-02-01 for an end-to-end simulation of underground muon-induced secondaries.

\begin{figure}[ht]
    \centering
    \includegraphics[width=\linewidth]{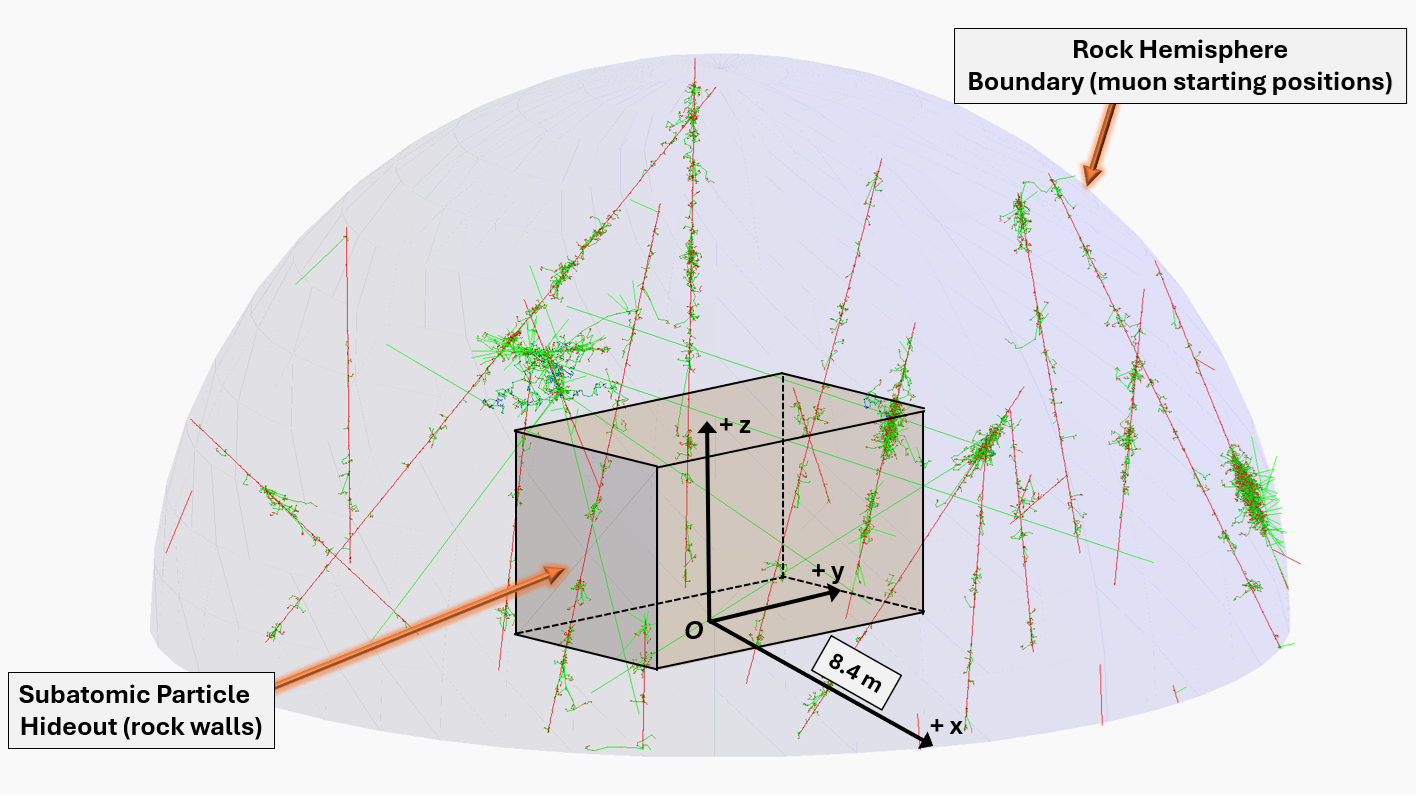}
    \caption{Reconstruction of the \textsc{geant4} geometry for the SPH. Red lines are primary muons and all others are muon-induced secondaries.}
    \label{sph}
\end{figure}

Our simulation strategy separates the problem into two distinct stages, each optimized for its specific physics requirements. In the first stage, \textsc{mute} simulates the transport of primary muons through a realistic overburden from the surface to the boundary of a hemispheric rock volume centered on the lab (illustrated in Fig.~\ref{sph}), yielding realistic muon energy and angular distributions. A hemispheric geometry is chosen because underground muons and their secondaries predominantly arrive from the upper half-hemisphere, making it the minimal volume that fully captures the relevant angular phase space. The rock hemisphere radii used for SPH and CLI are \SI{8.4}{m} and \SI{12.0}{m}, respectively. In the second stage, \textsc{geant4} takes these muon distributions as input and simulates the production and transport of secondary particles in the vicinity of the underground laboratory. This division allows us to leverage the strengths of each simulation toolkit while avoiding computationally prohibitive full-geometry simulations.

\textsc{mute} is a Python-based framework designed for computational efficiency, coupling \textsc{mceq}~\cite{mceq3} and \textsc{proposal}~\cite{proposal} to handle muon propagation through flat and complex overburden geometries. It allows the user to define propagation materials, select atmospheric and hadronic interaction models, and handles both flat and mountain topologies. For this work, \textsc{mute} generates site-specific underground muon energy and angular spectra by propagating the primary cosmic-ray muon flux through the detailed overburden profiles of each underground location.

\textsc{geant4} is a simulation toolkit for particle-matter interactions. It provides physics processes for electromagnetic, hadronic, and optical interactions, along with geometry definition and material specification capabilities. In our approach, \textsc{geant4} is tasked with transporting the muon flux generated by \textsc{mute} through a model of the underground laboratory and surrounding rock. The simulation includes muon energy loss, multiple scattering, and the full cascade of secondary particles produced via muon-nuclear interactions, photonuclear reactions, and electromagnetic showers.

The justification for this two-stage approach lies in the physics of secondary particle production in thick absorbers. After sufficient muon track length in rock, the production of secondaries and their lateral spread relative to the muon direction approaches equilibrium between generation and absorption. Once this equilibrium is reached, the probability of a secondary escaping the rock into the laboratory depends only on the local rock composition and thickness, not on the full propagation history from the surface. This shower equilibrium principle allows us to approximate the complex overburden surrounding the laboratory with a simplified geometry in \textsc{geant4} while maintaining accuracy in secondary particle yields and escape probabilities.

In the \textsc{geant4} stage, each underground laboratory is modeled as a cavern filled with air surrounded by a homogeneous hemisphere of rock with composition and density representative of the local geology. The muon flux from \textsc{mute}, carrying the correct energy and angular distributions for each specific site, is injected at the surface of the outer boundary of this rock volume. All muon-induced secondaries are tracked through the rock, and their relevant observables are recorded when they cross the boundary delineating the rock and the cavern walls, referred to as the rock-cavern boundary, to extract the fluxes and energy distributions of particles entering the laboratory. An example of simulated geometries for the SPH is shown in Figure \ref{sph}, where the red lines correspond to primary muons and all others are muon-induced secondaries.

Secondaries are tagged and terminated upon their first entry from the rock into the cavern, so back-scattering from air back into the rock is not modeled. This approximation is appropriate because the goal of this stage is to define the first-entry flux incident on the laboratory boundary; any scattering or moderation after entry is treated in downstream detector simulations, and incorporating it here would double count those processes and distort the low-energy flux. The flux at the rock-cavern boundary is given by

\begin{equation}
    \Phi_{\rm RC} = \frac{N_{\rm s}}{T_{\rm live}\times A},
\end{equation}

where $N_{\rm s}$ is the number of secondaries, $T_{\rm live}$ is the simulation live time, and $A$ is the area of the lab walls which excludes the floor as upgoing muons are not considered. The areas used for SPH and Cryolab I are \SI{66.2}{\square\metre} and \SI{211.8}{\square\metre}, respectively.

The secondary underground fluxes, differential energy spectra, and angular intensities are computed from the simulation counts \(N_{ijk}\), where \(i\), \(j\), and \(k\) index the zenith-angle \(\theta_i\), azimuth-angle \(\phi_j\), and energy-bin \(E_k\), respectively. The solid-angle width of each angular bin is 
\(\Delta\Omega_{ij} = \Delta\phi\,(\cos\theta_i - \cos\theta_{i+1})\), the energy-bin width is \(\Delta E_k = E_{k+1}-E_k\), and the angle-integrated counts in each energy bin are 
\(N_k = \sum_{i=0}^{N_\theta-1}\sum_{j=0}^{M_\phi-1} N_{ijk}\), where \(N_\theta\) and \(M_\phi\) denote the number of zenith and azimuth bins.

\begin{equation}
\frac{d\Phi}{dE\,d\Omega}
=
\frac{N_{ijk}}{A\,T_{\rm live}\,\Delta\Omega_{ij}\,\Delta E_k}
\;[\si{\per\square\metre\per\second\per\steradian\per\mega\electronvolt}].
\label{diff_fluxes}
\end{equation}

\begin{equation}
\frac{d\Phi}{dE}
=
\frac{N_{k}}{A\,T_{\rm live}\,\Delta E_k}
\;[\si{\per\square\metre\per\second\per\mega\electronvolt}].
\label{diff_spectrum}
\end{equation}

\begin{equation}
\frac{d\Phi}{d\Omega}
=
\frac{\sum_{k=0}^{N_{E}-1} N_{ijk}}{A\,T_{\rm live}\,\Delta\Omega_{ij}}
\;[\si{\per\square\metre\per\second\per\steradian}].
\label{intensities}
\end{equation}

For the full-scale simulations, 400 million muons were generated directly at the radius of the rock-hemisphere boundary for each underground location, corresponding to live times of \SI{30.5}{days} and \SI{15.6}{days} for the SPH and CLI sites, respectively. These live times were obtained by first counting the number of primary muons crossing a reference area of \SI{1}{\square\metre} at the origin ($N_{\rm hits}$), (0,0,0), of each underground location. This \SI{1}{\square\metre} patch was chosen for simplicity, as fluxes are conventionally expressed per unit area (e.g., per cm$^2$ or m$^2$), allowing a straightforward conversion from the simulated muon counts to an equivalent real-time exposure by normalizing to the total muon flux at that site ($\Phi_{\mu}$), via $T_{\rm live}$ = $N_{\rm hits}\text{ / }\Phi_{\mu}$. The difference in live times between the two locations arises because simulating CLI in \textsc{geant4} requires a larger volume, which lowers the relative fraction of simulated muons contributing to the effective flux.

For SPH, the measured flux of \SI{0.239}{} \(\pm  \text{ }0.025_{\mathrm{sys}}\)\,\si{\per\square\metre\per\second} was used, while for CLI, the simulated flux of \SI{0.259}{} \(\pm\text{ }0.026_{\mathrm{sys}}\)\,\si{\per\square\metre\per\second} was used. These fluxes are taken from previous work~\cite{dakota} and the systematic errors result in uncertainties in the live times. The uncertainty in the lab areas comes from the nonuniformity of the walls, which effectively changes the mean free path of the secondaries. Since this is nontrivial to quantify, a conservative error of $\pm$ \SI{4}{\square\metre} and $\pm$ \SI{13}{\square\metre} for SPH and CLI, respectively, is applied. The systematic uncertainties in live times ($\sigma_{T_{live}}$) and laboratory areas ($\sigma_{\rm A}$) are detailed in Table \ref{systematics}. 

\subsection{Neutron Production in \textsc{geant4}}
\textsc{geant4} offers a wide range of physics lists. For the simulation of neutrons below $<$ \SI{20}{MeV}, relevant for accurate underground and shielding studies, \textsc{geant4} provides access to the neutron High Precision (HP) package. The neutron-HP package uses precomputed data from Evaluated Nuclear Data Files (ENDF), and is available as an extension to several existing physics lists. Two widely accepted packages used for neutron production simulations \textsc{ftfp\_bert\_hp}~\cite{ALLISON2016186} and \textsc{shielding\_hp}~\cite{geant4_Shielding} (analogous to \textsc{shielding}) are compared. A comprehensive comparison of neutron production in rock was established by running the full-scale simulation using the \textsc{ftfp\_bert\_hp} and \textsc{shielding\_hp} lists. 

For this comparison, we looked at all of the muon-induced neutrons which crossed from the rock volume into the lab space for the SPH. The observables we compared were the neutron creation processes, the total neutron flux at the rock-cavern boundary, the neutron exit energies as they crossed the boundary, and the distance at which the neutrons were produced from the lab walls. The creation processes are the direct processes taken when a neutron is first created. The neutron exit energies are the energies at the point where they exit from the rock into the lab. The production distances are the straight-line distances from the creation coordinate to the position at which the neutrons enter the lab.

\begin{figure}[ht]
  \centering
  \begin{subfigure}[b]{1.0\linewidth}
    \centering
    \includegraphics[width=\linewidth]{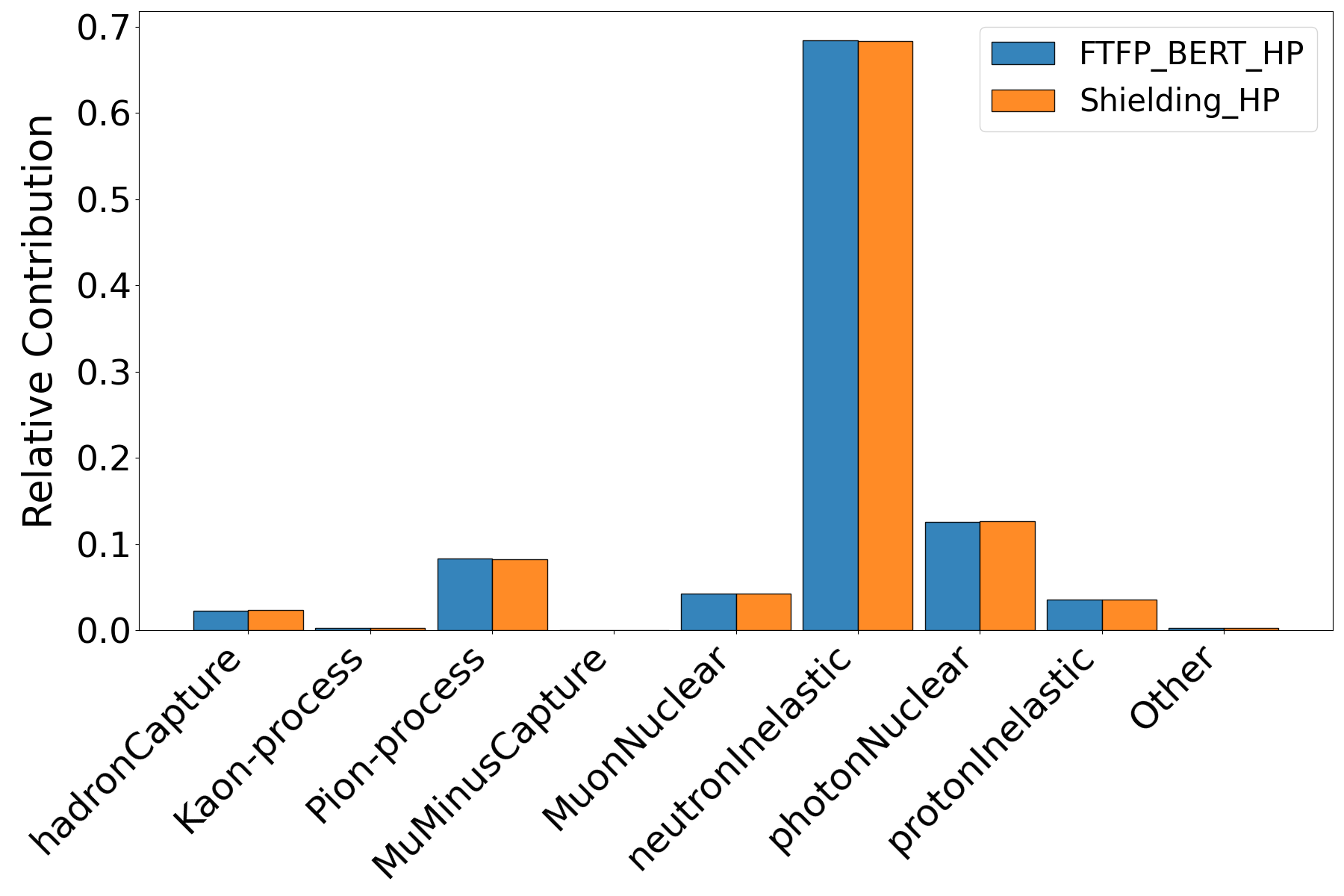}
    \caption{Relative contribution of neutron creation processes between \textsc{ftfp\_bert\_hp} and \textsc{shielding\_hp}.}
    \label{fig:creation_proc}
  \end{subfigure}

  \vspace{6pt} 

  \begin{subfigure}[b]{1.0\linewidth}
    \centering
    \includegraphics[width=\linewidth]{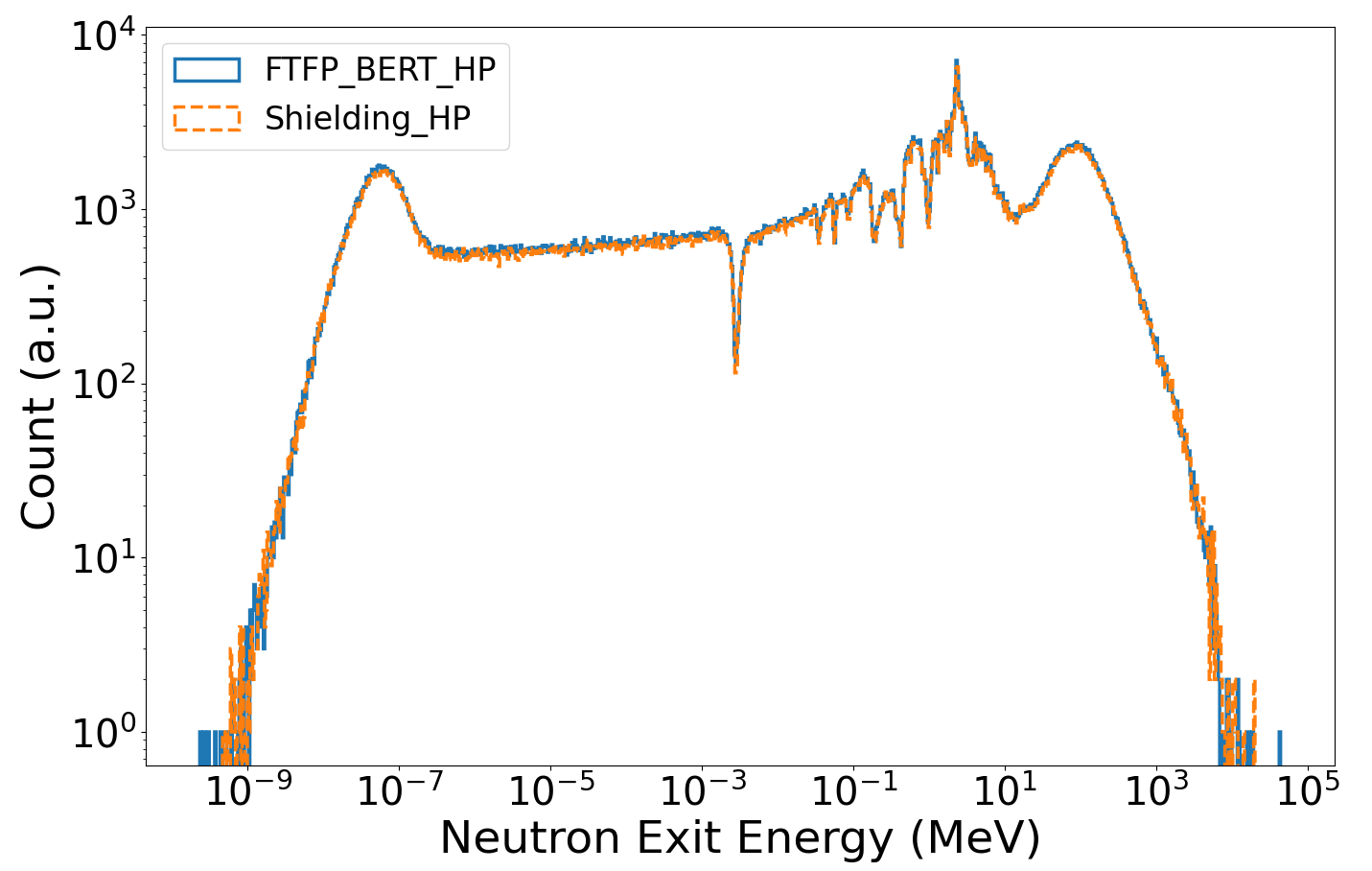}
    \caption{Comparison of the neutron exit energies, crossing from the rock into the cavern, between \textsc{ftfp\_bert\_hp} and \textsc{shielding\_hp}.}
    \label{fig:phys_list_energies}
  \end{subfigure}

  \vspace{6pt}

  \begin{subfigure}[b]{1.0\linewidth}
    \centering
    \includegraphics[width=\linewidth]{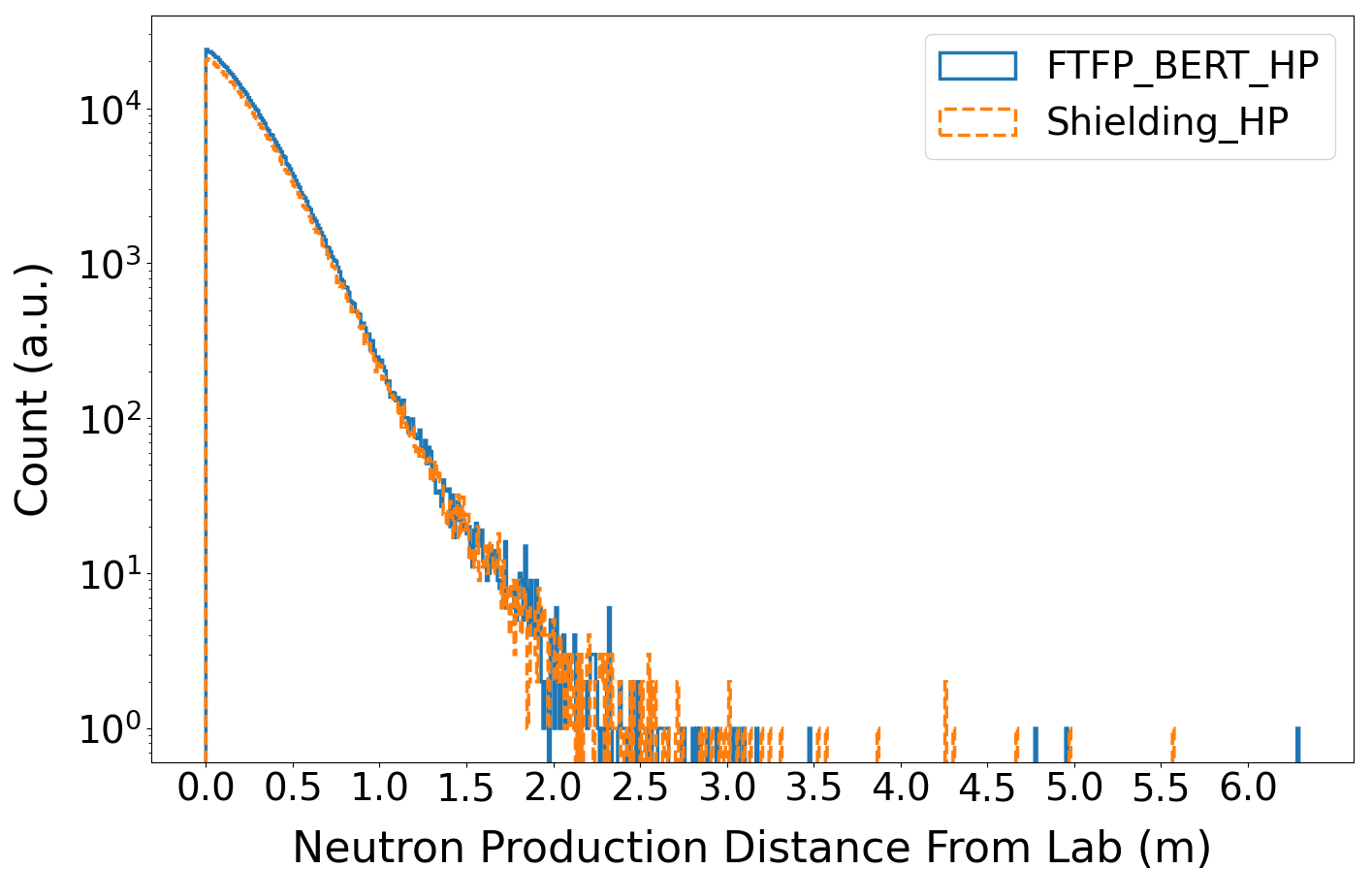}
    \caption{Comparison of the straight-line neutron production distance between \textsc{ftfp\_bert\_hp} and \textsc{shielding\_hp}.}
    \label{fig:phys_list_dist}
  \end{subfigure}

  \caption{Comparisons between \textsc{ftfp\_bert\_hp} and \textsc{shielding\_hp}:
    (a) neutron creation processes,
    (b) neutron exit energies, and
    (c) neutron production distances.}
  \label{phys_list_combined}
\end{figure}

Figures~\ref{phys_list_combined} shows the differences between the \textsc{ftfp\_bert\_hp} and \textsc{shielding\_hp} physics lists for the neutron creation processes, exit energies, and production distances. The deviation between \textsc{ftfp\_bert\_hp} and \textsc{shielding\_hp} is shown to be negligible, with a difference in the final neutron flux of $<$ \SI{1}{\%}. Therefore, since \textsc{shielding\_hp} was specifically developed for low-background simulations, this package was chosen for the final analysis, and the difference between the two models was taken as a systematic error ($\sigma_{\textsc{geant4}}$).

\subsection{Neutron Production Correction}

Previous studies have demonstrated that general-purpose Monte Carlo toolkits such as \textsc{fluka} and \textsc{geant4} can yield systematically different predictions for muon-induced neutron production, and both frameworks can underpredict total neutron yields relative to experimental measurements in certain regimes~\cite{ARAUJO2005398, REICHHART201367, g4_fluka_n_prod, sutanto_mu_n}. Detailed inter-comparisons show that, for muons with energies above $\sim$ \SI{100}{GeV}, \textsc{geant4} tends to produce fewer neutrons than \textsc{fluka}, attributed to differences in hadronic cascade modeling and secondary neutron multiplicities~\cite{ARAUJO2005398}. In comparisons to experimental data, \textsc{geant4} predictions are often lower by factors of 2 or more, particularly in thick, low-Z targets and high-Z materials such as lead, with discrepancies attributable to limitations in the treatment of photo-nuclear and muon-nuclear processes as well as hadronic shower development~\cite{abe_mu_n, KNEIBL201987, MARINO2007611}. These differences have important implications for background estimates in rare-event searches and motivate continued refinements of neutron production models in Monte Carlo codes.

To quantitatively assess the performance of the specific \textsc{geant4} version used in this work, we simulated the muon-induced neutron yield in several liquid-scintillator compositions reported in Refs.~\cite{abe_mu_n, aberdeen, hertenberger_mu_n, boehm_mu_n, Enikeev1987, zhang_mu_n} over a range of mean underground muon energies spanning \SIrange{13}{340}{GeV}. The simulated yields reproduce the correct energy dependence but systematically fall below the experimental measurements, by factors ranging from approximately 1.4 to 2.8, except at the lowest energies where agreement is better. This persistent underproduction across a wide range of materials and energies indicates that the discrepancy is not dominated by detector- or material-specific effects, but instead reflects a broader shortfall in the underlying neutron-production modeling.

\begin{figure}[ht]
    \centering
    \includegraphics[width=\linewidth]{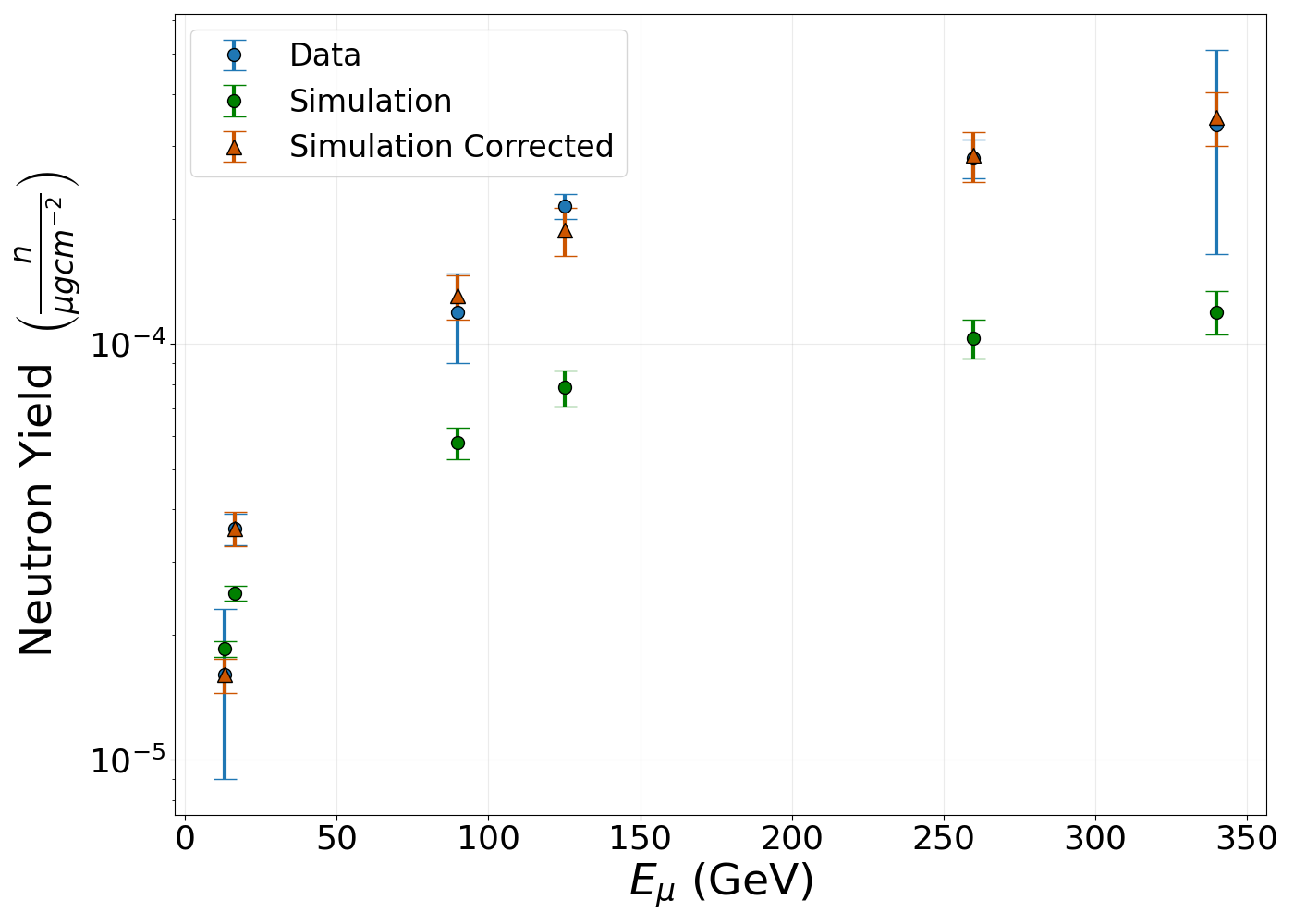}
    \caption{Comparison of the simulated liquid scintillator neutron yields, before and after applying the correction, against the experimental data from Refs.~\cite{abe_mu_n, aberdeen, hertenberger_mu_n, boehm_mu_n, Enikeev1987, zhang_mu_n}}
    \label{sim_exp_yields_corrected}
\end{figure}

Therefore, following the methodology introduced by Mei \& Hime (MH) in Ref.~\cite{mei_hime}, we extract an empirical correction function to compensate for this bias. We determine the correction by fitting the relative difference between the measured neutron yields and our \textsc{geant4} predictions as a function of muon energy, leading to
\begin{equation}
M_{\rm corr}
= \frac{M_{\rm sim}}{1 - 0.314\,\cdot E_{\mu}^{0.128} + 1.68\times10^{6}\,\cdot E_{\mu}^{-5.793}}\, ,
\label{neutron_correction}
\end{equation}
where $M_{\rm sim.}$ is the simulated neutron multiplicity, $E_{\mu}$ is the muon energy, and $M_{\rm corr.}$ is the corrected multiplicity. The improvement obtained by applying Eq.~\ref{neutron_correction} is illustrated in Fig.~\ref{sim_exp_yields_corrected}. This correction is applied to all \textsc{geant4} neutron results reported in this work. The uncertainty introduced by the correction varies by only a few percent between the lowest and highest muon energies considered; therefore, we adopt a conservative, energy-independent uncertainty of \SI{8.5}{\%} ($\sigma_{\rm corr}$) for all corrected neutron fluxes.


\section{Underground Muons}\label{sec:MuonSpec}
The muon-induced secondary production largely depends on the underground muon energy distribution and the composition of the local overhead rock. The material through which a muon propagates dictates the amount and type of interactions; more specifically, the muon cross section for ionization is proportional to the average proton number, $\langle Z\rangle$, spallation events are proportional to the average nucleon number, $\langle A\rangle$, and pair production and Bremsstrahlung are proportional to the ratio of $\langle Z^{2}/A\rangle$ \cite{Kokoulin1971ICRC12, Kelner1999PAN, Kelner1995MuonBrems, Abramowicz1991PLB, Abramowicz1997ALLM, Rossi1952HEP}.

To ensure an accurate modeling of the overburden rock, several rock samples were taken from both the Subatomic Particle Hideout and Cryolab I. Scanning electron microscopy (SEM) was performed on each sample by the Colorado School of Mines Department of Geology \cite{mines_sem_website}. The SEM method offers a powerful and straightforward analysis of mineral characterization \cite{sem}. The results of this analysis are shown in Table \ref{rock_comp} and a custom rock composition, corresponding to these values, was constructed independently in \textsc{geant4} for the SPH and CLI.

\vspace{0.25cm}
\begin{table*}[ht!]
\footnotesize
\centering
\renewcommand{\arraystretch}{1.5}
\setlength{\tabcolsep}{8.0pt}
\caption{Chemical composition of the major components, given in mass fraction ($\%$), density, and the $\langle Z\rangle$, $\langle A\rangle$, and $\langle Z^{2}/A\rangle$ values for the Subatomic Particle Hideout and Cryolab I.}

\begin{tabular}{l|c|c|c|c|c|c|c|c|c|c|c|c}
\toprule
\multicolumn{13}{c}{} \\

\textbf{Location}
& Si
& O
& Al
& K 
& Na
& Mg
& Fe
& Other
& Density [\unit[per-mode=symbol]{\gram\per\cubic\centi\metre}]
& $\langle Z\rangle$
& $\langle A\rangle$
& $\langle Z^{2}/A\rangle$\\

SPH
& 36.22 
& 49.84  
& 5.89  
& 1.65  
& 3.47 
& 1.08
& 1.24  
& 0.61 
& 2.69 $\pm$ 0.06
& 11.06
& 22.27
& 5.49 \\

Cryolab I
&  36.01
&  49.71
&  5.47
&  2.40
&  2.42
&  1.84
&  1.17
& 0.98 
& 2.71 $\pm$ 0.07
& 11.13
& 22.41
& 5.53 \\

\end{tabular}

\label{rock_comp}
\end{table*}

The underground muon energy spectrum also plays an important role in the secondary production rate. The muon-induced neutron production is often approximated by using the mean underground muon energy and is reported to follow a power-law with $E_{\mu}^{-n}$, where $E_{\mu}$ is the mean muon energy and $n$ ranges from 0.7-0.8 \cite{mei_hime, PhysRevD.64.013012, LINDOTE2009366}. However, this approximation ignores complex physics processes, as well as angular effects from the overburden, and can lead to discrepancies with experimental results of order \SI{25}{\%} \cite{REICHHART201367}. Obtaining the underground muon energy spectrum through experimental efforts is not a trivial process, and a secondary method relies on estimating the full spectrum through parameterizations of the underground muon flux as a function of energy, zenith angle, and depth, such as those discussed in Refs. \cite{Reichenbacher2008Crouch, mei_hime, bugaev_spectra, taba_spec, daya_bay_spec}. Although this sampling method provides a more faithful muon energy distribution to the true underground distribution, it also has the effect of averaging out angular dependencies arising from topological anisotropy, in particular, azimuthal anisotropy. This is less of an issue for facilities under a flat surface since the azimuth muon distribution is assumed to be isotropic, but for those with complex overburden profiles, like CURIE, there is a non-negligible effect. For this reason, we have chosen to develop angular-dependent muon energy sampling. 

Previous work on the underground muon flux at CURIE has validated \textsc{mute} as an accurate and reliable simulation framework~\cite{dakota}. Therefore, to construct our site-specific underground muon energy spectra for the simulations, we use the corresponding differential fluxes derived from \textsc{mute}, $\Phi_{\mu}(E_{\mu}, \theta, \phi)$, where $E_{\mu}$ is the underground muon energy and ($\theta$, $\phi$) is the zenith and azimuth muon arrival angles, respectively. For a given arrival direction, there is a corresponding energy distribution, $(E_{\mu}, I_{\mu})$, where $I_{\mu}$ is the intensity associated with each underground energy. For each arrival angle, we integrate the energy distribution over the intensities to obtain an angle-dependent mean muon energy. Thus, we construct an energy sampling grid with the same resolution as the arrival angles. Although this is still an approximation, it has the advantage over sampling the full angle-integrated energy spectrum by means of both zenith- and azimuth-dependent energy sampling.


To test the difference between a mean energy approximation and sampling from the angle-dependent muon energy spectrum for all muon-induced secondaries, simulations were also done using the \textsc{mute} derived mean energy for the Subatomic Particle Hideout, where $\langle E_{\mu}\rangle$ = \SI{98}{GeV}\footnote{The mean underground muon energy derived from \textsc{mute} for CLI is also \SI{98}{GeV}.}. When directly comparing the secondary production of $\langle E_{\mu}\rangle$ and the full energy spectrum, we note that on average the muon-induced secondary fluxes are higher for $\langle E_{\mu}\rangle$ than for the sampled energy spectrum, albeit within error. Table \ref{full_v_mean_flux} shows a direct comparison for the neutron and electromagnetic component ($\gamma$-ray, $e^{-}$, and $e^{+}$) fluxes simulated at the Subatomic Particle Hideout. For each secondary, the mean energy results in a higher flux absolute flux. Furthermore, the neutron yield derived from the $\langle E_{\mu}\rangle$ simulations gives an absolute difference of \SI{7.6}{\%} relative to the neutron flux from the full spectrum sampling. This discrepancy is in agreement with previous work \cite{Empl_2014}, and highlights the need for proper sampling.

\begin{figure}[ht]
    \centering
    \includegraphics[width=\linewidth]{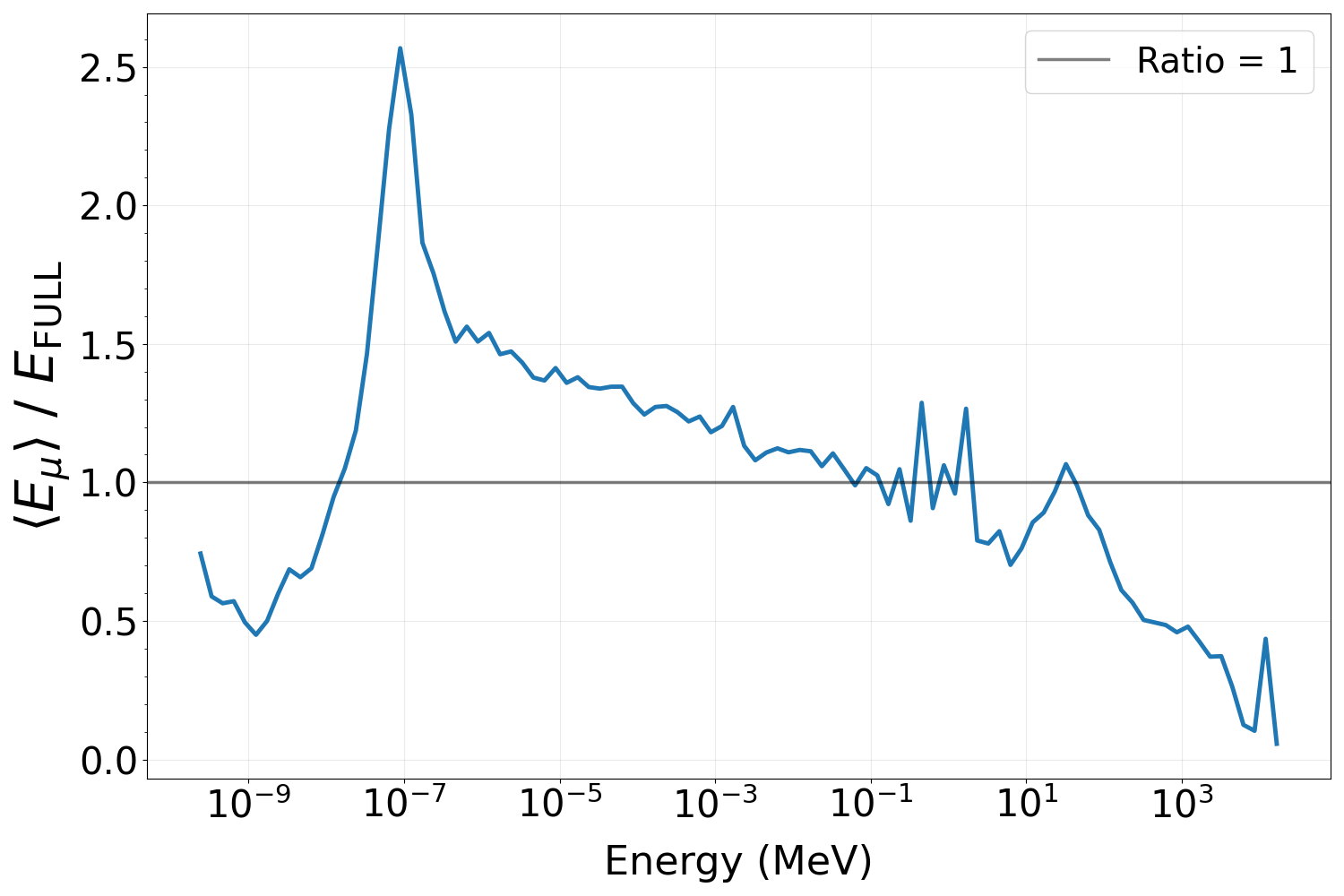}
    \caption{Comparison of the spectral ratios between $\langle E_{\mu}\rangle$ and $E_{\rm FULL}$.}
    \label{mean_v_full_ratio}
\end{figure}

Furthermore, by comparing the ratio of the mean to full neutron spectral ratios, $\langle E_{\mu}\rangle$/$E_{\rm FULL}$, as a function of energy (Fig.~\ref{mean_v_full_ratio}), it is clear that the mean-energy approximation does not reproduce the correct spectral behavior. The ratio departs noticeably from unity across most energies, exceeding a factor of two at both low and high secondary energies and remaining systematically offset at intermediate energies. The underlying energy-dependent discrepancies demonstrate that the mean-energy approximation fails to capture the correct spectral shape, further reinforcing the need for full sampling of the muon energy distribution in precision background studies.

\vspace{0.25cm}
\begin{table}[ht]
\footnotesize
\centering
\renewcommand{\arraystretch}{1.35}
\setlength{\tabcolsep}{5.5pt}
\caption{Comparison of the secondary flux between mean energy and full muon spectrum sampling for the Subatomic Particle Hideout.}

\begin{tabular}{l|c|c}
\textbf{Secondary} & \textbf{$\Phi_{FULL}$} [\si{\per\square\metre\per\second}] & \textbf{$\Phi_{\langle E_{\mu}\rangle}$} [\si{\per\square\metre\per\second}] \\ \hline
Neutron  & 
(8.52 $\pm$ 1.30$_{\text{sys.}}$)$\times10^{-3}$ & 
(9.17 $\pm$ 1.40$_{\text{sys.}}$)$\times10^{-3}$ \\
 Gamma   & 
 (5.54 $\pm$ 0.70$_{sys.}$)$\times10^{-1}$ & 
(5.82 $\pm$ 0.74$_{sys.}$) $\times10^{-1}$ \\
 Electron  &
 (4.39 $\pm$ 0.56$_{sys.}$)$\times10^{-2}$ &
(4.52 $\pm$ 0.57$_{sys.}$)$\times10^{-2}$ \\
Positron & 
(1.08 $\pm$ 0.14$_{sys.}$)$\times10^{-2}$ &
(1.15 $\pm$ 0.15$_{sys.}$)$\times10^{-2}$\\
\end{tabular}

\label{full_v_mean_flux}
\end{table}

\subsection{Shower Equilibrium}
While the shower equilibrium principle provides the theoretical justification for our simplified rock geometry in \textsc{geant4}, the minimum rock thickness required to achieve this equilibrium (constant average secondary production per muon) must be empirically determined for the specific material composition at each site. To establish this critical thickness, we conducted a systematic study using a rectangular rock volume with dimensions \SI{15}{m} $\times$ \SI{10}{m} $\times$ \SI{10}{m} simulated in \textsc{geant4}.

Muons were injected along the long axis of the volume, and the number of secondary particles and their transverse propagation distances (perpendicular to the muon momentum) were recorded in \SI{0.5}{cm} segments from the surface down to a depth of \SI{10}{m}. Special attention was given to neutron counting to prevent double-counting from inelastic scattering events, where \textsc{geant4} terminates the original neutron track and registers all resulting final-state neutrons as new particles. This simulation was repeated using the full energy spectrum ($E_{\rm FULL}$) for both SPH and CLI sites, as well as monoenergetic beams at $\langle E_{\mu}\rangle$ = \SI{98}{GeV}, \SI{50}{GeV}, \SI{500}{GeV}, and \SI{5000}{GeV} to assess energy-dependent equilibrium behavior.

\begin{figure}[ht]
    \centering
    \includegraphics[width=\linewidth]{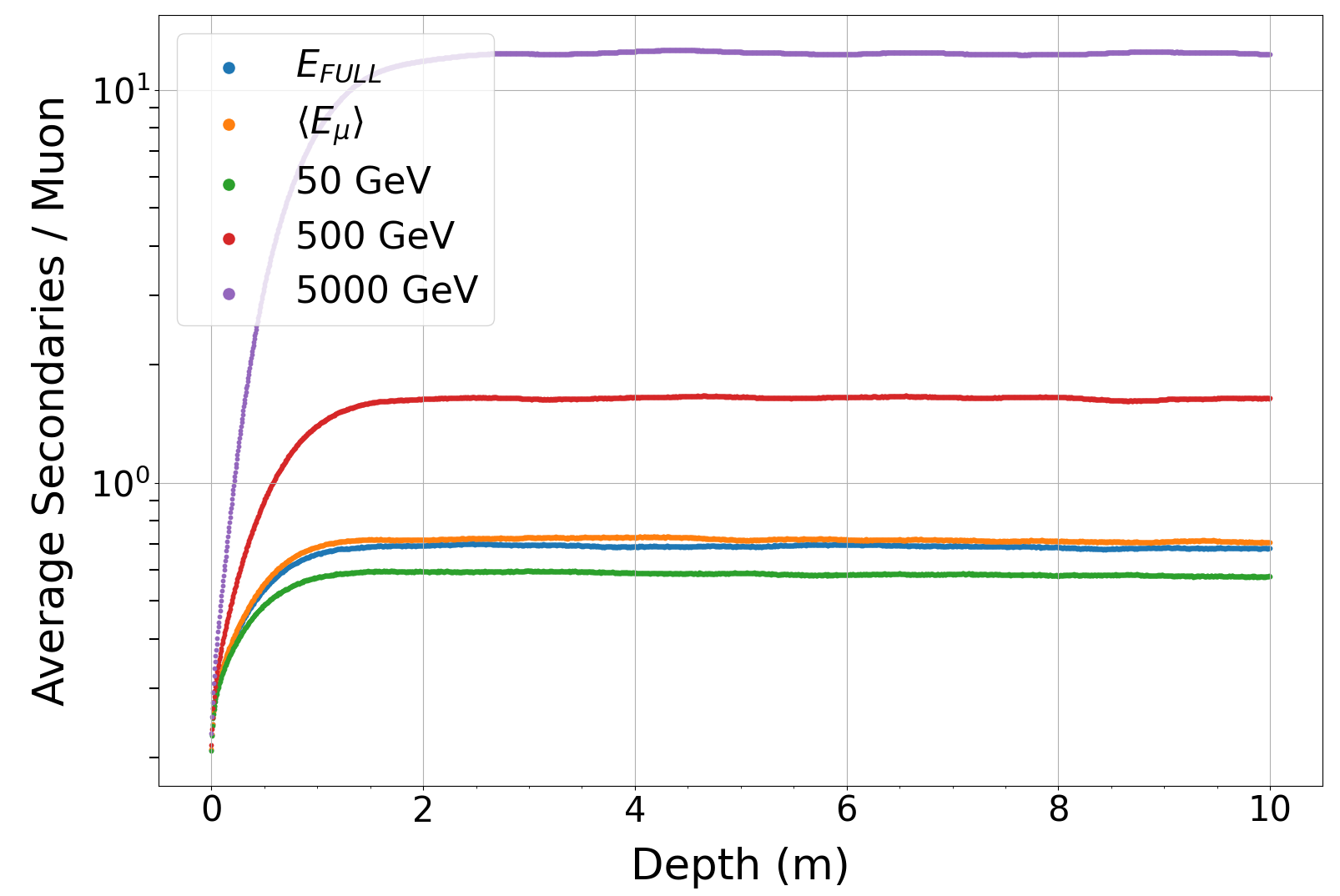}
    \caption{Average secondary production number per muon as a function of depth.}
    \label{avg_sec_prod}
\end{figure}
\begin{figure}[ht]
    \centering
    \includegraphics[width=\linewidth]{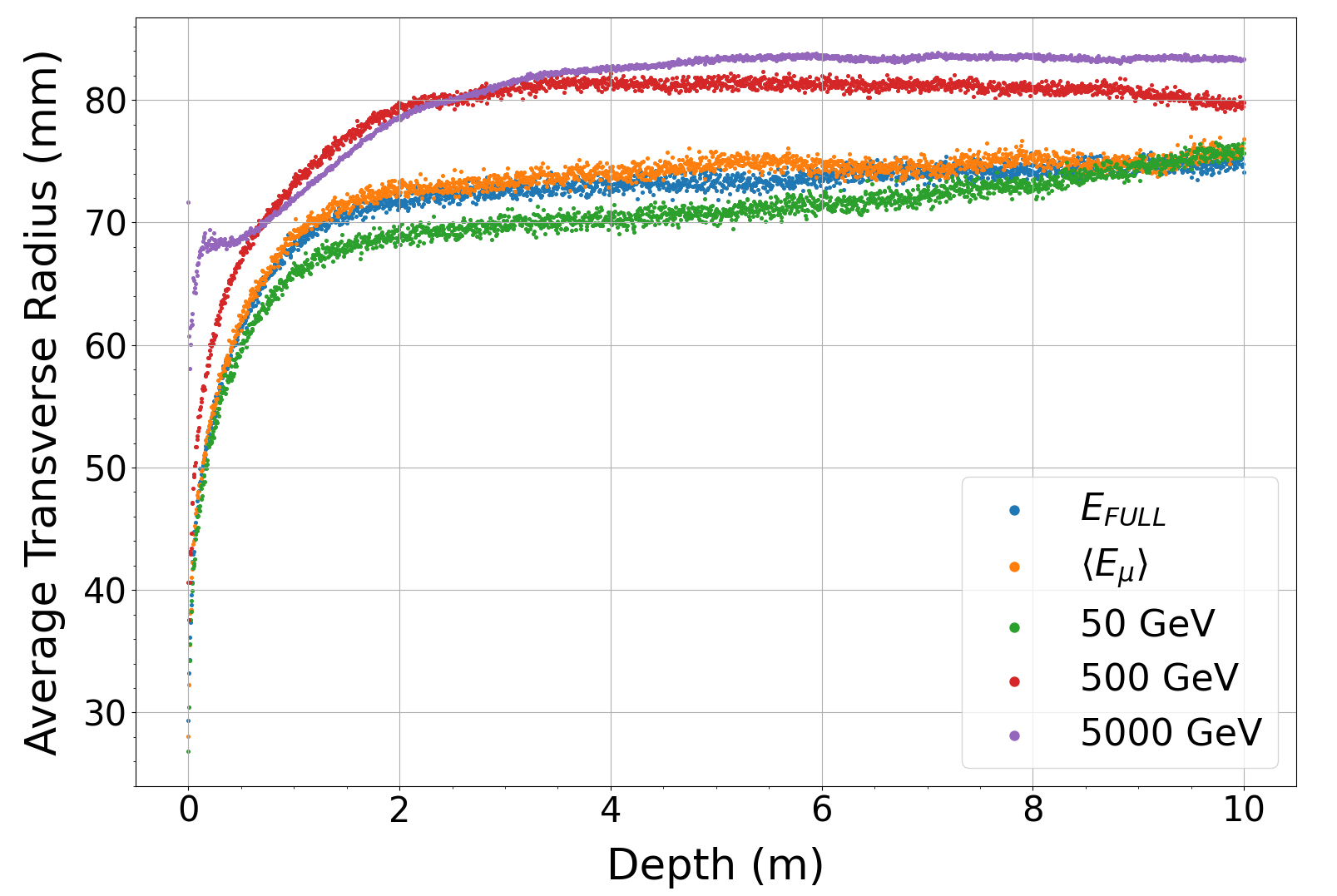}
    \caption{Average secondary transverse propagation distance as a function of depth.}
    \label{avg_trans}
\end{figure}

The simulation results for the SPH are shown in Figures \ref{avg_sec_prod} and \ref{avg_trans}, where the transverse distance and secondary production exhibit a slight non-zero slope beyond equilibrium, which arises from muon energy loss in the rock during propagation. With respect to the transverse distance, this corresponds to a relatively higher amount of secondary transverse momentum; however, the variation in this change, on the order of of a few mm, is much less than the average secondary transverse propagation distance at stability. Therefore, this effect does not result in any significant change in the total number of secondaries capable of escaping the rock into the cavern, and thus no systematic uncertainty is assigned.

\begin{figure}[ht]
    \centering
    \includegraphics[width=\linewidth]{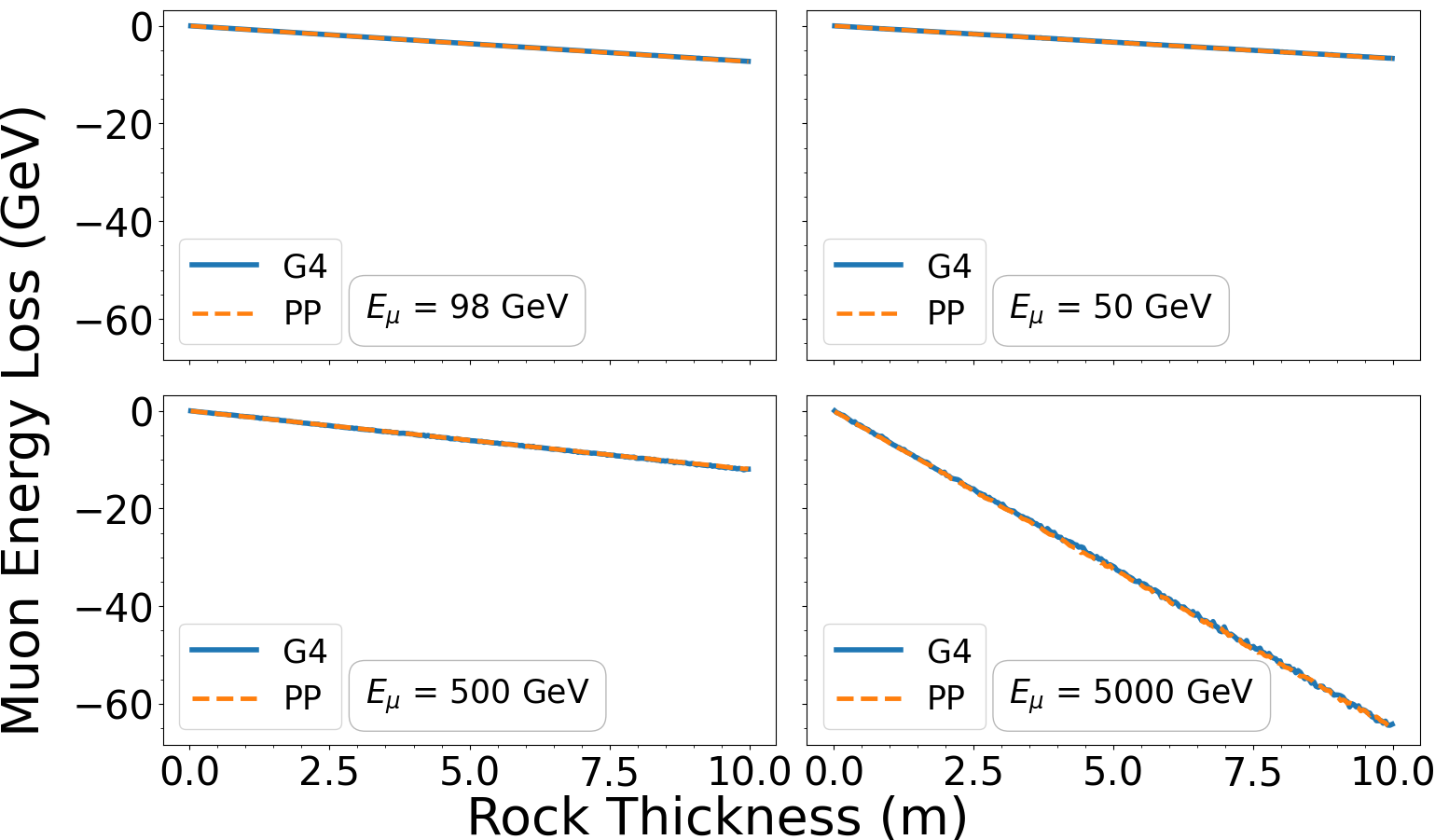}
    \caption{Direct comparison of muon energy loss in rock between \textsc{proposal} (PP) and \textsc{geant4} (G4).}
    \label{pp_v_g4}
\end{figure}

The non-zero slope in the secondary production is expected, since muon energy loss is required to produce secondary particles. The concern would be large differences in muon energy loss when transferring primary muons from \textsc{mute} to \textsc{geant4} to complete the propagation to depth. Performing a direct comparison of the muon energy loss through \SI{10}{m} of rock between \textsc{proposal} and \textsc{geant4} shows a negligible difference between the muon energies tested (see Fig.~\ref{pp_v_g4}). Furthermore, since the SPH and CLI geometries correspond to minimum and maximum depths that are only subsets of the total \SI{10}{m}, any discrepancy in energy loss is necessarily smaller than that demonstrated for the full depth and therefore negligible for the purposes of this analysis. Because the comparison establishes that the potential differences between the two propagation methods are below the level of relevance to the underground flux determination ($\sim$ \SI{1e-2}{\%}), we assign no additional systematic uncertainty associated with this effect.

Based on these results, the transverse propagation is found to reach equilibrium at approximately \SI{4}{m} from the face of the rock volume, while secondary production stabilizes at about \SI{2}{m}. The results from the CLI energy spectrum and rock composition show no statistical difference. Furthermore, we find an independence of the equilibrium depths with the muon energy, in agreement with previous findings \cite{cao2024cosmicrayinducedneutron}. From these results, we select a minimum rock thickness of \SI{4}{m} and construct a site-specific hemisphere of rock in \textsc{geant4}, such that the minimum muon propagation distance from the surface of the hemisphere to any point on the lab wall is \SI{4}{m}. Moreover, by selecting a minimum thickness of \SI{3}{m} or \SI{5}{m} instead, the average secondary production number is changed across all tested muon energies by less than 0.4\%, and because the escaping flux is directly related to the local production, such variations are negligible for our analysis.

\begin{figure}[ht]
    \centering
    \includegraphics[width=\linewidth]{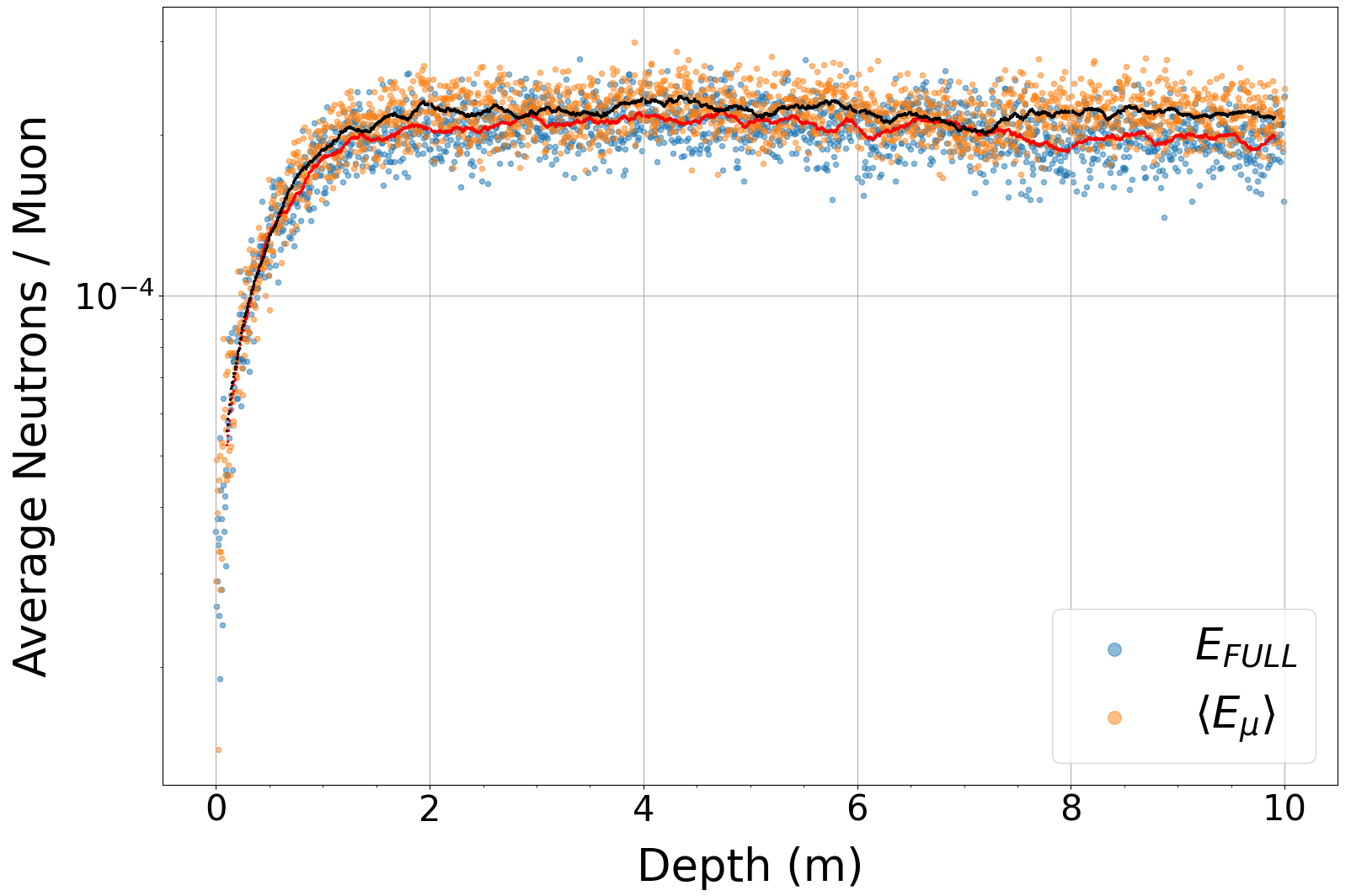}
    \caption{Average neutron production number per muon as a function of depth for $\langle E_{\mu}\rangle$ (orange) and the full spectrum sampling (blue). Individual points show the raw results for 0.5 cm rock segments. The solid lines, $\langle E_{\mu}\rangle$ (black) and the full spectrum sampling (red), represent 20 cm running averages, included only to highlight the underlying systematic trend and aid visualization; all quantitative results are based on the raw 0.5 cm data.}
    \label{neutron_prod}
\end{figure}

To further investigate the effect of approximating the energy spectrum with $\langle E_{\mu}\rangle$, we compared the two methods directly in the equilibrium study. Figure \ref{neutron_prod} shows the neutron production as a function of depth, where agreement is seen only in the first meter of rock, prior to saturation. Beyond this point, the two functions diverge and the relative difference across all depths is \SI{7.1}{\%} and \SI{7.6}{\%} beyond saturation, in agreement with the result from the previous section.

%

\subsection{Spectra Corrections}
\textsc{mute} treats the labs as point-source locations, and therefore the total rock through which a muon propagates is inflated by the average distance from the center of the lab to the walls. Before the muon angular and energy distributions were used in the full-scale simulations, a correction was needed to account for the finite size of underground labs and the simulated hemisphere of rock discussed in the previous sections. To make this correction, the area-weighted average distance from the center of the SPH and CLI dimensions was calculated and \SI{4}{m} was added to this to account for the hemisphere of rock. This distance was converted to a slant depth in kilometer-water-equivalent, subtracted from the site-specific mountain profiles discussed in \cite{dakota}, and the \textsc{mute} simulations were repeated to obtain the corrected muon energy and angular distributions. 


\subsection{Muon Bundles}
Underground muons arrive either as single events or in bundles (multiplicity greater than one), with the relative contributions of each providing sensitivity to both the energy and composition of the primary cosmic rays in the atmosphere \cite{Petkov:2011tda}. The muon multiplicity distribution has been extensively investigated in deep underground laboratories, where single-muon events account for approximately \SI{94}-\SI{97}{\%} of the total flux \cite{macro_bundle, macro_bundle2, frejus_bundle, lvd_bundle, soudan_bundle, bust_bundle}. In contrast, far fewer measurements exist at shallower depths relevant to facilities such as CURIE. The limited studies available in the \SI{100}-\SI{320} m.w.e.\ range indicate that the single-muon fraction increases to $\gtrsim$ \SI{97}{\%} \cite{Tcaciuc:2006zz, emma_bundle, Shtejer_2016}.

To quantify the potential influence of these rare multi-muon events on secondary particle production, we adopted the multiplicity fractions reported by Tcaciuc \textit{et al.}~\cite{Tcaciuc:2006zz}, whose measurement depth ($\sim$ \SI{320} m.w.e.) is closest to that of CURIE. Using these fractions, we performed an approximate numerical study in which simulated events were randomly reassigned to multiplicity classes (\textit{M} = 1–4) and their secondary yields scaled accordingly; in all cases, the resulting shifts in particle fluxes were found to be small, at the level of $\lesssim$ 0.5\%. Based on this, and for simplicity, we treat muons as statistically uncorrelated and consider only single-muon events in the present work.

\subsection{Systematic Errors}
All uncertainties presented in this work are purely systematic, as they dominate over statistical errors. The largest systematic errors come from the uncertainty in the rock density ($\sigma_{\rho_{\rm rock}}$) and the muon energies ($\sigma_{E_{\mu}}$), which both alter the secondary production rate. To test the effect of these parameters, secondary production was analyzed during the equilibrium study by individually changing the rock density and muon energies, as well as investigating combined changes. The result of the combined uncertainty changes on the average secondary production versus depth, where the largest deviations from the baseline $E_{\rm FULL}$ simulations are $+\sigma_{\rho_{\rm rock}}\text{ /}+\sigma_{E_{\mu}}$ and $-\sigma_{\rho_{\rm rock}}\text{ /}-\sigma_{E_{\mu}}$. We note here that the density and energy effects are positively correlated and are assigned a single systematic uncertainty.

Table \ref{systematics} lists all systematics included and their absolute and relative contributions to the final uncertainties in the secondary fluxes. Because Sternheimer parameters, which govern the density correction in the Bethe–Bloch equation, are not known for most laboratory rock types, calculations must adopt values from standard rock. Woodley \textit{et al.} have shown that this approximation leads to a nearly constant 10\% shift in the predicted underground muon intensities~\cite{woodley2024cosmicraymuonslaboratories}, which is independent of depth. Following their recommendation, we therefore assign an additional \SI{10}{\%} systematic uncertainty ($\sigma_{\rm SH}$) to the MUTE-predicted muon flux for CLI, which is subsequently propagated through the secondary production simulations. The SPH muon flux is experimentally measured and therefore does not get assigned this additional uncertainty.



\vspace{0.25cm}
\begin{table}[hb]
\footnotesize
\centering
\renewcommand{\arraystretch}{1.35}
\setlength{\tabcolsep}{5.5pt}
\caption{Systematic errors used in this study and their contributions to the total secondary flux.}

\begin{tabular}{l|c|c}
\textbf{Systematic} & \textbf{Absolute} & \textbf{\% $\Phi_{\rm RC}$} \\ \hline
 $\sigma_{\rm corr.}$ & -- & $\pm$ \SI{8.5}{} \\
 
 $\sigma_{\rho_{\rm rock}}$ [SPH(CLI)] & $\pm$ 0.06(0.07) [\si{\gram\per\cubic\centi\metre}] & -- \\
 
 $\sigma_{E_{\mu}}$   & $\pm$ \SI{3}{\%} [\si{\giga\electronvolt}] & -- \\
 
 $\sigma_{\rho_{\rm rock}}$ + $\sigma_{E_{\mu}}$   & -- & $\pm$ \SI{5}{} \\

  $\sigma_{\rm SH}$ [CLI]  & -- & $\pm$ \SI{10}{} \\
 
 $\sigma_{\rm T_{live}}$ [SPH(CLI)] &  $^{\text{+}3.56}_{\text{--}2.88}$($^{\text{+}1.73}_{\text{--}1.42}$) [\si{\day}] &  $\pm$ \SI{10}{} \\

 $\sigma_{\rm A}$ [SPH(CLI)] & $\pm$ \SI{4.0}(\SI{13.0}) [\si{\square\metre}] &  $\pm$ \SI{6.0}{} \\
 
  $\sigma_{\rm \textsc{geant4}}$ & -- & $\pm$ \SI{0.5}{} \\
\end{tabular}

\label{systematics}
\end{table}

\subsection{Depth-Intensity Relation}
Ref.~\cite{mei_hime} provides a convenient \textsc{fluka}-based parameterization of the muon-induced neutron flux at the rock-cavern boundary as a function of the facility depth. As noted earlier, the neutron fluxes presented in this depth parameterization have been inflated using a muon-induced neutron multiplicity correction, in order to better represent experimental data. Extrapolating this relationship to the depth of CURIE, the predicted value disagrees within 1$\sigma$ of the values derived for SPH and CLI in this work. For further comparison, we have also simulated the neutron flux for Gran Sasso, Soudan, Kamioka, and the Sudbury Neutrino Observatory Laboratory (SNOLAB) using the rock compositions provided in Ref.~\cite{woodley2024cosmicraymuonslaboratories} and the methodology detailed in the previous sections, where we find good agreement with the MH model.

\begin{figure}[ht]
    \centering
    \includegraphics[width=\linewidth]{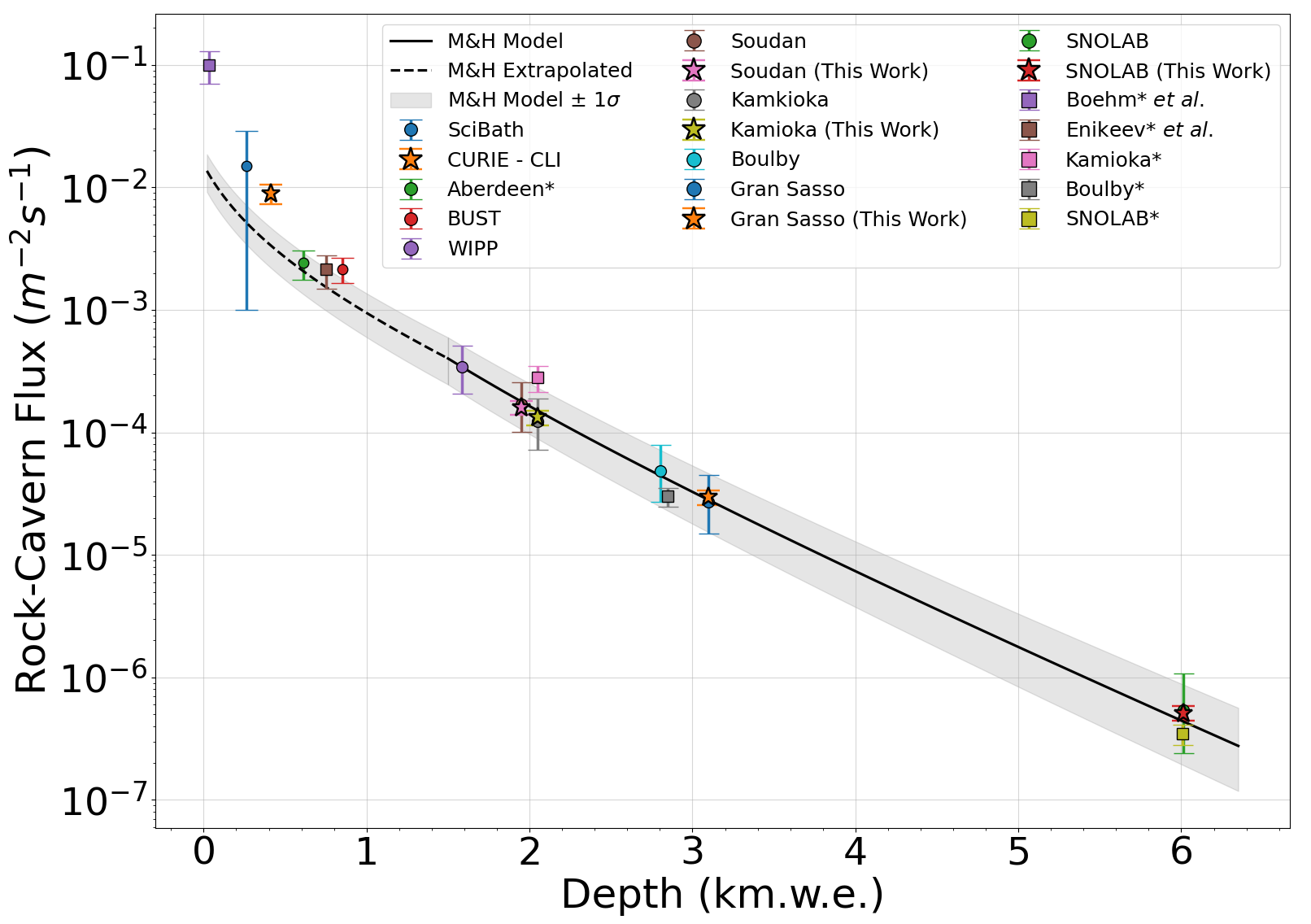}
    \caption{Muon-induced neutron flux parameterization from Ref.~\cite{mei_hime}. Also shown are the neutron fluxes for SciBath~\cite{Garrison:2014nla} and BUST~\cite{bust_n_flux}, those estimated (denoted with *) from~\cite{aberdeen, abe_mu_n, zhang_mu_n, Enikeev1987, boehm_mu_n, ARAUJO2008471, collaboration2025cosmogenicneutronproductionwater}, and the fluxes from this work. The errors shown are taken from reported experimental uncertainties, except for those predicted from this work, which are purely systematic.}
    \label{mh_n_flux}
\end{figure}

Figure \ref{mh_n_flux} shows the parameterization and the predicted fluxes taken directly from \cite{mei_hime}, along with the multiplicity corrected simulated fluxes from this work, as well as the experimental fluxes from SciBath \cite{Garrison:2014nla} and the Baksan Underground Scintillation Telescope (BUST) \cite{bust_n_flux}. Using the experimental data and design information provided in Refs.~\cite{aberdeen, abe_mu_n, zhang_mu_n, Enikeev1987, boehm_mu_n, ARAUJO2008471, collaboration2025cosmogenicneutronproductionwater}, we have also converted the muon-induced neutron yields in these works to an approximate muon-induced neutron flux. The muon-induced neutron flux of \SI{21.5e-5}{n\per\square\metre\per\second} presented in \cite{bust_n_flux} is likely misquoted as an order of magnitude lower. The parameterization from \cite{mei_hime} used to validate this result yields \SI{15.1e-4}{n\per\square\metre\per\second}, which is an order of magnitude higher than quoted in~\cite{bust_n_flux}. The discrepancy likely arises from a unit or normalization mismatch in the original value.



\begin{figure}[ht]
    \centering
    \includegraphics[width=\linewidth]{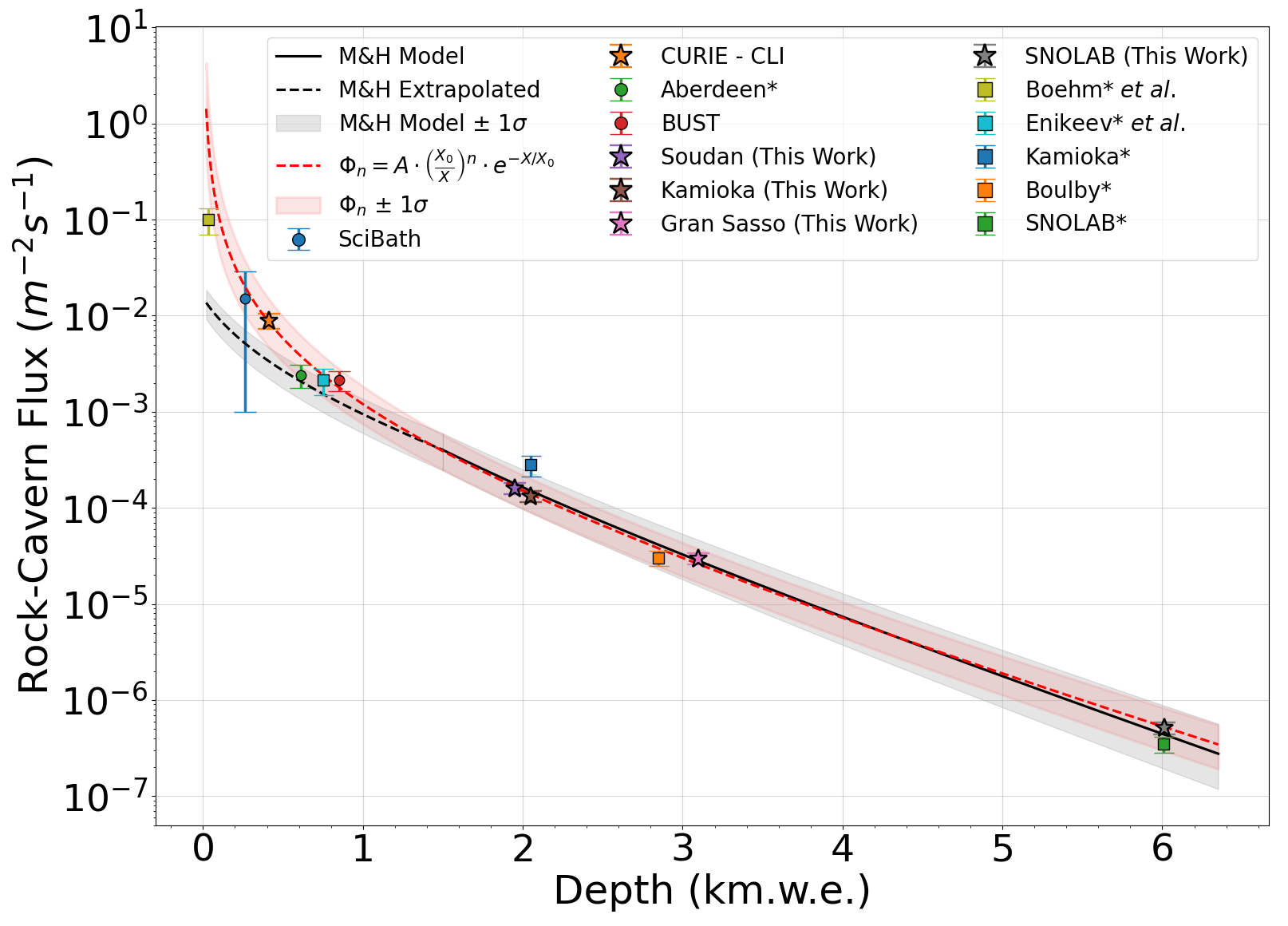}
    \caption{Muon-induced neutron flux parameterization from Eqn.~\ref{eq:kmwe_model} applied to the simulated fluxes from this work, compared to the MH model. Also shown are the neutron fluxes for SciBath~\cite{Garrison:2014nla} and BUST~\cite{bust_n_flux}, and those estimated (denoted with *) from~\cite{aberdeen, abe_mu_n, zhang_mu_n, Enikeev1987, boehm_mu_n, ARAUJO2008471, collaboration2025cosmogenicneutronproductionwater}. The errors shown are taken from reported experimental uncertainties where available, except for those predicted from this work, which are purely systematic.}
    \label{mh_n_flux_new}
\end{figure}

Agreement is seen between the extrapolation of the MH model and the neutron fluxes above $\sim$ \SI{0.6} km.w.e\,. However, below this, there is a clear deviation of the model predictions. To avoid these extrapolations and to provide a more robust model for both shallow- and deep-underground facilities, we present a new depth-intensity relation (DIR) by fitting the simulated fluxes presented in this work:

\begin{equation}
    \Phi_{\rm n}(\Chi) = A\cdot \left(\frac{\Chi_{0}}{\Chi}\right)^{n}\cdot e^{-\Chi/\Chi_{0}}\,,
    \label{eq:kmwe_model}
\end{equation}
where $A = \SI{0.003}{} \pm \SI{0.001}{\per\square\metre\per\second}$, $\Chi_{0} = \SI{1.015}{} \pm \SI{0.083}{} \text{ km.w.e.}$, and $n = \SI{1.562}{} \pm \SI{0.192}{}$. The results of this fitting procedure, compared to the MH model, can be seen in Fig.~\ref{mh_n_flux_new}. Furthermore, a measurement of the fast-neutron flux (\(E_{\rm n} > \SI{20}{MeV}\)) from muon-induced events was performed at the Soudan Underground Laboratory, yielding \((2.23 \pm 0.52_{\rm stat} \pm 0.99_{\rm sys}) \times 10^{-5}\,\si{\per\square\metre\per\second}\)~\cite{zhang_soudan}. This is in good agreement with the prediction from our simulation framework of \((2.32\, \pm\, 0.314_{\rm sys}) \times 10^{-5}\,\si{\per\square\metre\per\second}\).

\section{Results}
 As stated earlier, the major components of concern from the muon-induced backgrounds are neutrons and the electromagnetic component. Even though muon-induced electron and muon (anti)neutrinos have fluxes similar to that of the neutrons, neutrinos rarely interact and the solar neutrino flux is on the order of \SI{1e14}{\per\square\metre\per\second}~\cite{neutrino_flux}, making the ratio of the solar neutrino to muon-induced neutrino flux at CURIE $\approx$ 17 orders of magnitude higher. Therefore, we do not discuss them here as a background (see Table \ref{all_particle_fluxes}).

\subsection{Muon-induced Neutrons}

\vspace{0.25cm}
\begin{table}[ht]
\footnotesize
\centering
\renewcommand{\arraystretch}{1.35}
\setlength{\tabcolsep}{5.5pt}
\caption{Simulated muon-induced neutron fluxes over the full energy range and for different energy cuts.}

\begin{tabular}{l|c|c}
\textbf{Secondary} & \textbf{SPH} [\si{\per\square\metre\per\second}] & \textbf{Cryolab I} [\si{\per\square\metre\per\second}] \\ \hline
Neutron (all)               & (8.52 $\pm$ 1.30$_{\text{sys.}}$)$\times10^{-3}$ & (8.86 $\pm$ 1.62$_{\text{sys.}}$)$\times10^{-3}$ \\
 $>$ \SI{1}{MeV}   & (3.65 $\pm$ 0.56$_{\text{sys.}}$)$\times10^{-3}$ & (4.00 $\pm$ 0.73$_{\text{sys.}}$)$\times10^{-3}$ \\
 $>$ \SI{10}{MeV}  & (1.88 $\pm$ 0.29$_{\text{sys.}}$)$\times10^{-3}$ & (2.10 $\pm$ 0.38$_{\text{sys.}}$)$\times10^{-3}$ \\
 $>$ \SI{100}{MeV} & (7.77 $\pm$ 1.19$_{\text{sys.}}$)$\times10^{-4}$ & (8.75 $\pm$ 1.60$_{\text{sys.}}$)$\times10^{-4}$ \\
  $>$ \SI{1000}{MeV} & (3.32 $\pm$ 0.51$_{\text{sys.}}$)$\times10^{-5}$ & (3.74 $\pm$ 0.68$_{\text{sys.}}$)$\times10^{-5}$ \\
\end{tabular}

\label{neutron_flux}
\end{table}

\begin{figure}[ht]
    \centering
    \begin{subfigure}[b]{1.0\linewidth}
        \centering
        \includegraphics[width=\linewidth]{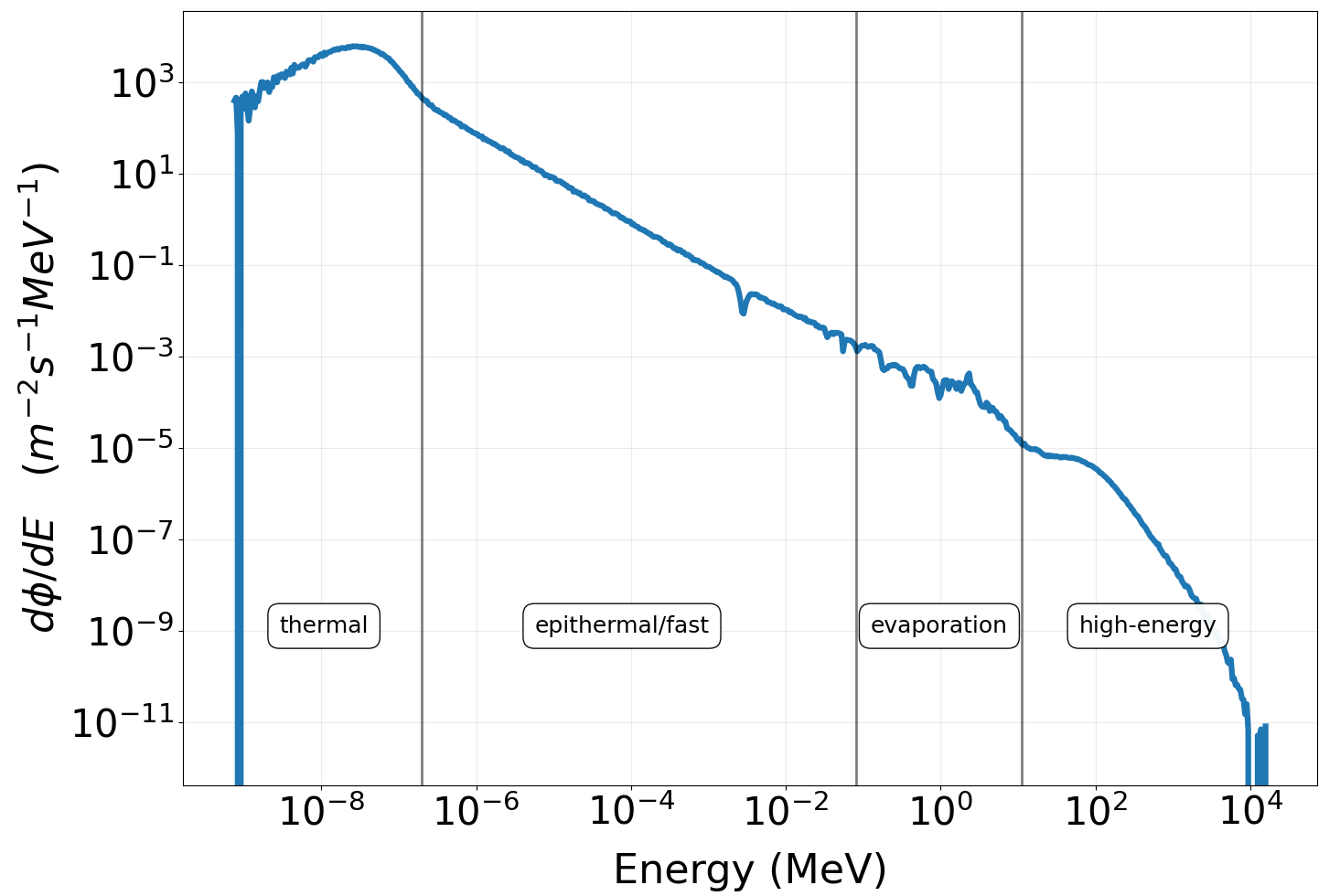}
        \caption{Muon-induced neutron energy spectrum at the rock-cavern boundary for the Subatomic Particle Hideout.}
        \label{n_energy_spec}
    \end{subfigure}
    \vspace{2em}

    \begin{subfigure}[b]{1.0\linewidth}
        \centering
        \includegraphics[width=\linewidth]{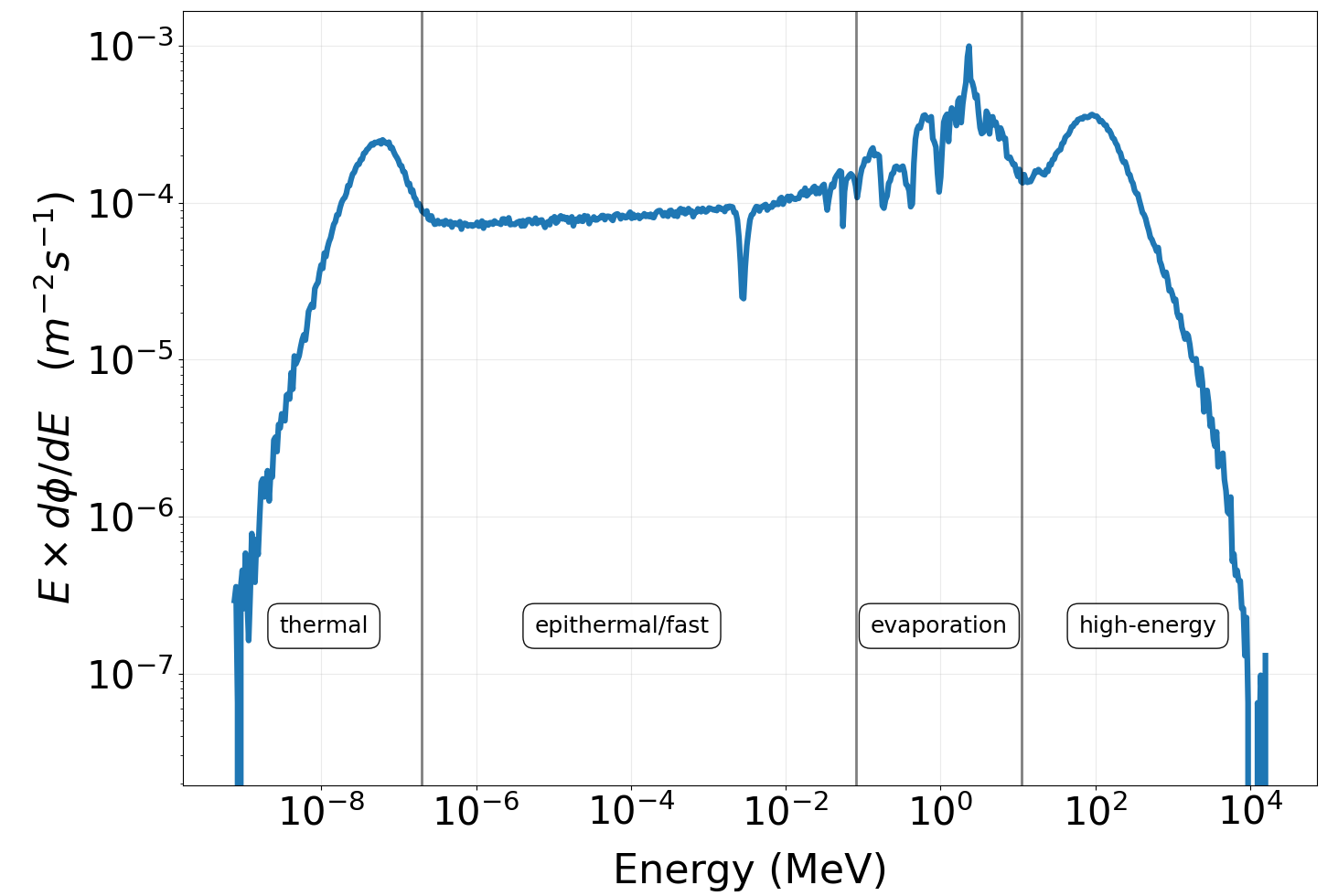}
        \caption{Energy-weighted muon-induced neutron energy spectrum at the rock-cavern boundary for the Subatomic Particle Hideout.}
        \label{n_scaled_spec}
    \end{subfigure}
    \caption{Neutron energy spectrum at the rock-cavern boundary of the Subatomic Particle Hideout: 
    (a) Unweighted
    (b) Energy-weighted. Although the transitions between regions are gradual, vertical lines are included to aid visual interpretation.}
    \label{neutron_energy_spectra}
\end{figure}

The muon-induced neutron rock-cavern fluxes are given in Table \ref{neutron_flux}, including the flux for several different energy cuts. The neutron energy spectrum for both locations is virtually identical, therefore only the SPH spectrum is shown in Figure \ref{neutron_energy_spectra}. In Figure \ref{n_scaled_spec}, each bin was weighted by the corresponding energy to emphasize the features of the neutron spectrum. This highlights an apparent absorption line just above $10^{-3}$ {MeV}. A plausible explanation for this feature is the small presence of $^{23}$Na in the composition of the rock (see Table~\ref{rock_comp}). The neutron capture cross sections on $^{23}$Na has a resonance precisely around the energy of this absorption, with a cross section of $\approx$ \SI{500}{mb} \cite{sodium}. Similar spectral features have been observed in simulations with NaCl overburden~\cite{dark_matter_neutron}, supporting this interpretation. However, other elements in the rock (e.g., Al~\cite{al_n_cap, al_n_cap2}, K~\cite{Macklin01101984}, and Fe~\cite{iron_n_capture}) can also contribute to capture in this energy region, making a definitive attribution nontrivial and beyond the scope this work.

\begin{figure}[htbp]
    \centering
    \begin{subfigure}[b]{1.0\linewidth}
        \centering
        \includegraphics[width=\linewidth]{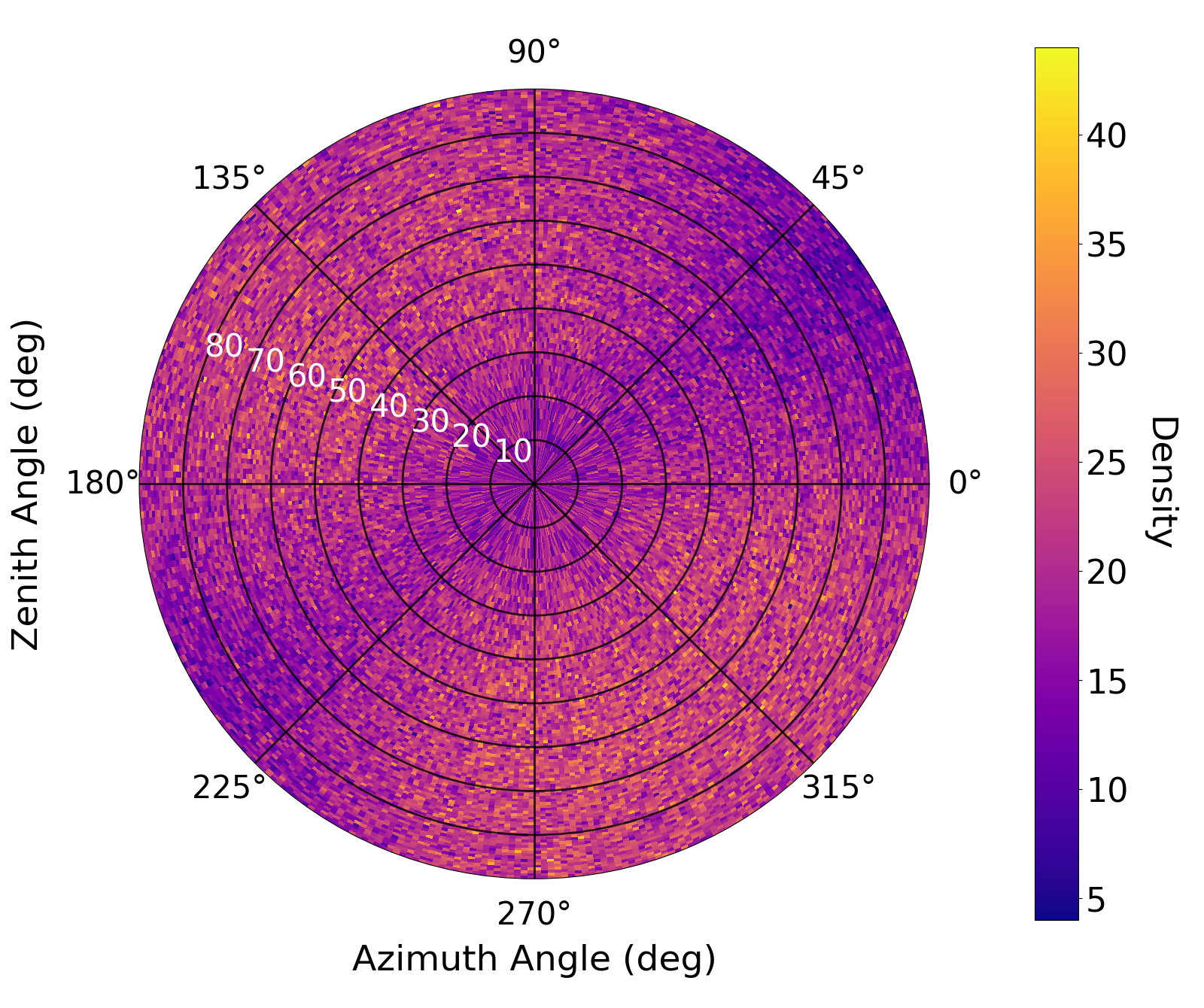}
        \caption{Neutron angular distribution at the rock-cavern boundary for the Subatomic Particle Hideout.}
        \label{fig:hideout_neutron_dist}
    \end{subfigure}

    \vspace{2em}

    \begin{subfigure}[b]{1.0\linewidth}
        \centering
        \includegraphics[width=\linewidth]{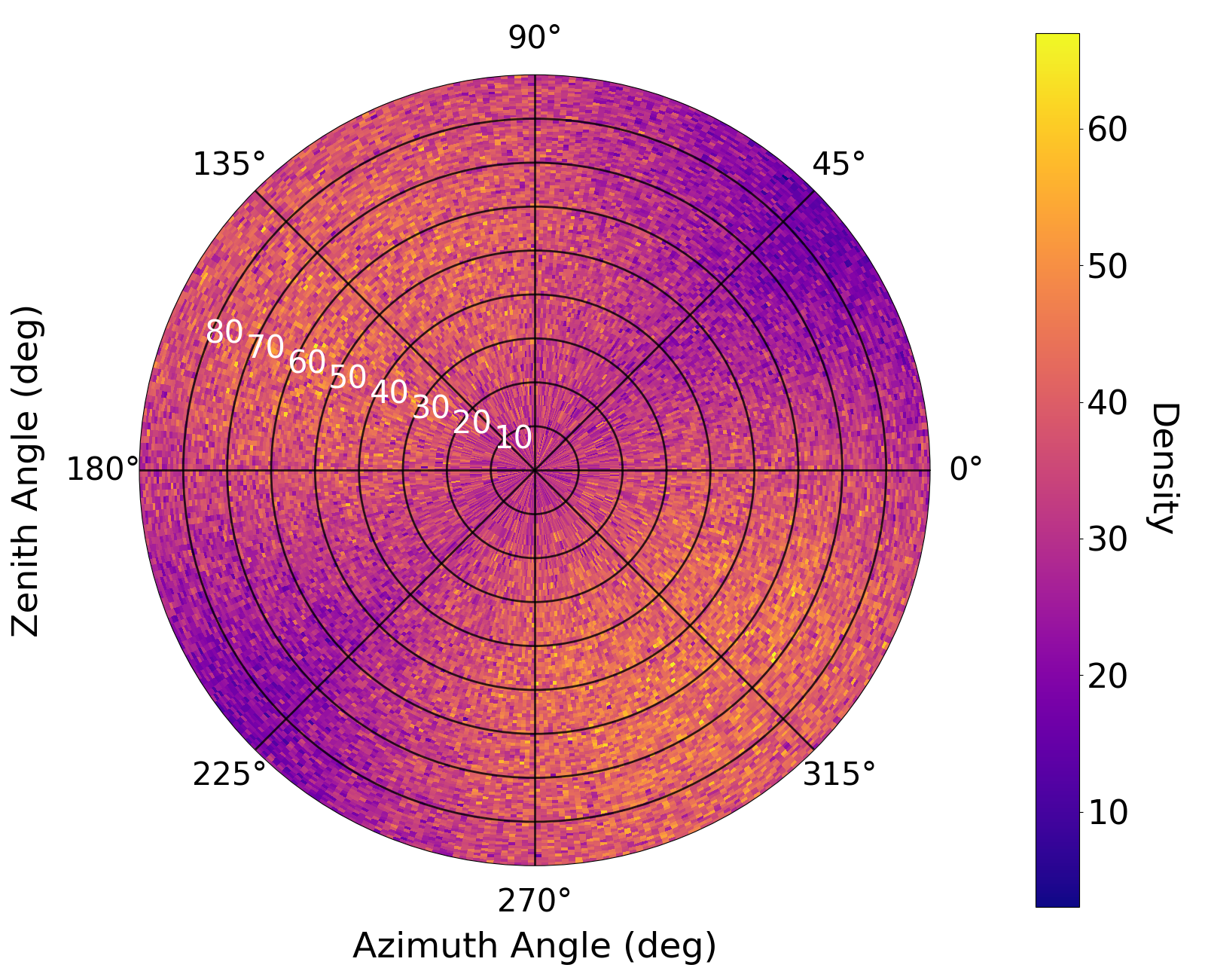}
        \caption{Neutron angular distribution at the rock-cavern boundary for Cryolab I.}
        \label{fig:cryolab_neutron_dist}
    \end{subfigure}

    \caption{Comparison of neutron angular distributions at the rock-cavern boundary: 
    (a) Subatomic Particle Hideout and 
    (b) Cryolab I. Note that \ang{90} corresponds to North in the compass quadrant system.}
    \label{neutron_dist_comparison}
\end{figure}

The neutron energy spectrum also shows four distinct regions: the thermal neutron peak around \SI{2.5e-8}{MeV}, the epithermal/fast region, the evaporation neutrons around \SI{0.1}{MeV} - \SI{10}{MeV}, and the high-energy spallation peak around \SI{100}{MeV}, with energies extending to several GeV. These spectral characteristics are consistent with previous simulation and experimental work \cite{cosmic_neutrons, felsenkeller_neutrons, KUDRYAVTSEV2003688, PhysRevC.72.025807}. Additionally, Figure \ref{neutron_dist_comparison} shows the angular distributions of neutrons entering the SPH and CLI and the rock-cavern boundary. These angular distributions are important because they determine the directions from which secondary neutrons illuminate the laboratory and therefore strongly influence shielding requirements and detector backgrounds.


\subsection{Electromagnetic Component}
\begin{figure}[ht]
    \centering
    \begin{subfigure}[b]{1.0\linewidth}
        \centering
        \includegraphics[width=\linewidth]{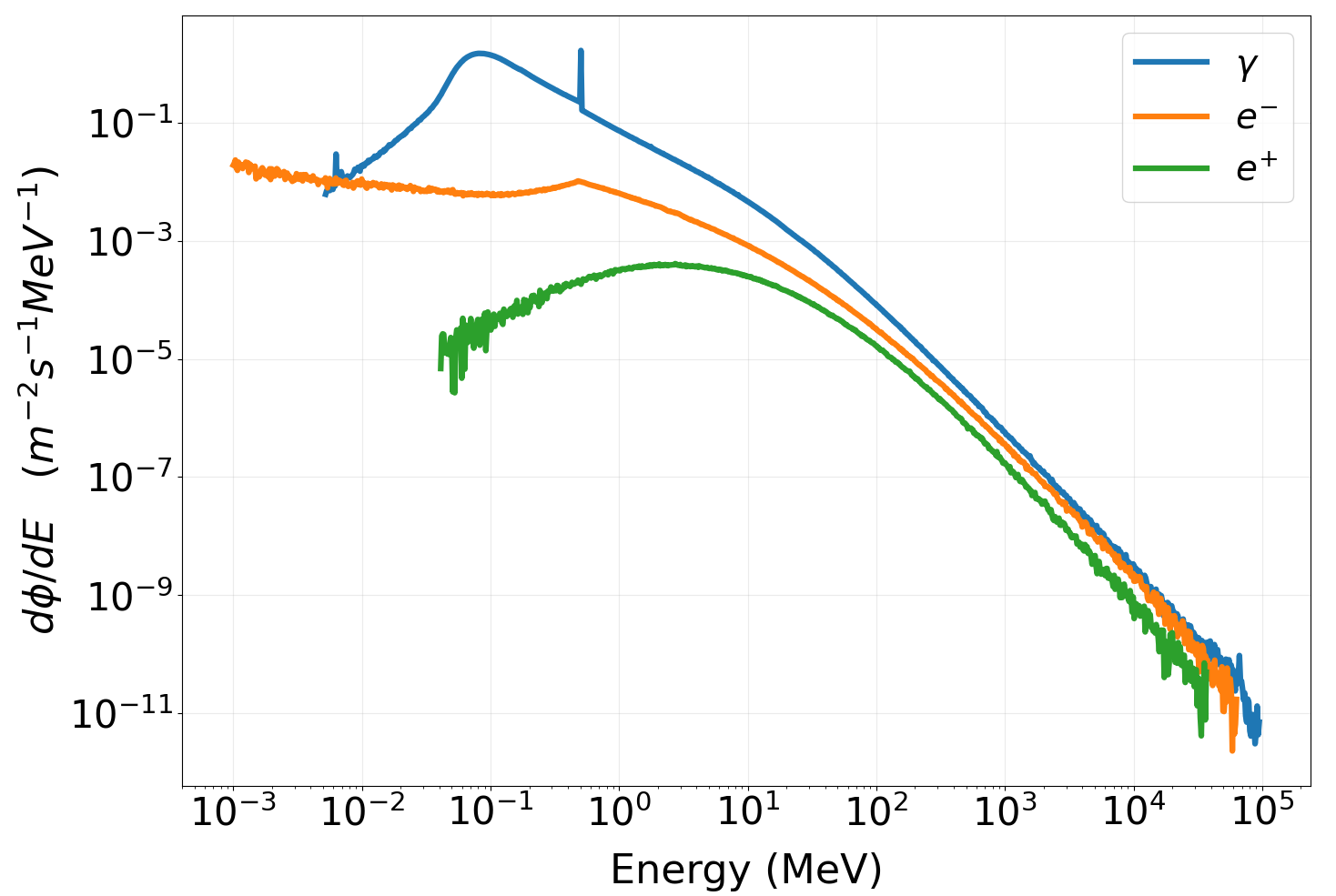}
        \caption{Muon-induced electromagnetic energy spectrum at the rock-cavern boundary for the Subatomic Particle Hideout.}
        \label{em_spec}
    \end{subfigure}
    \vspace{2em}

    \begin{subfigure}[b]{1.0\linewidth}
        \centering
        \includegraphics[width=\linewidth]{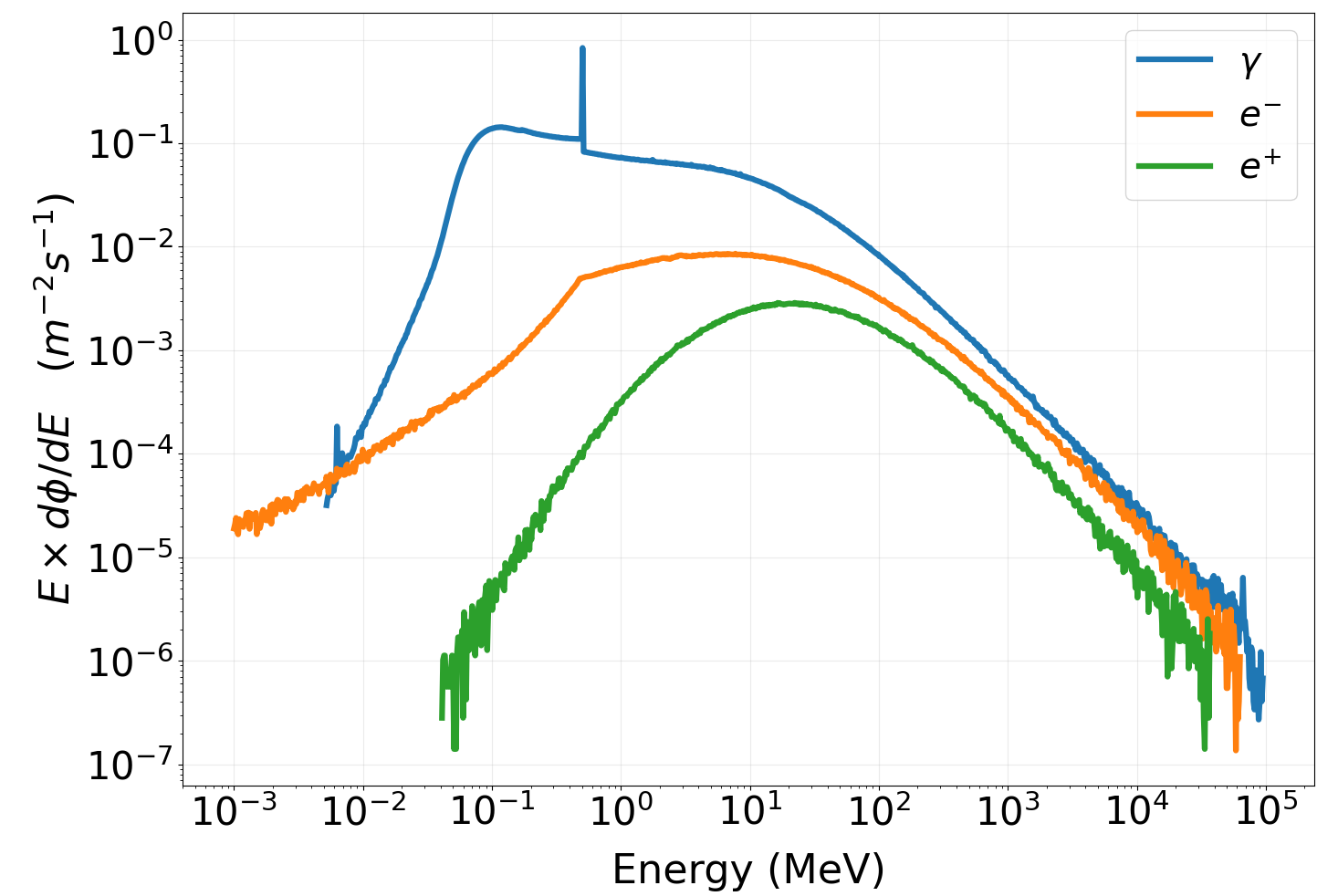}
        \caption{Energy-weighted muon-induced electromagnetic energy spectrum at the rock-cavern boundary for the Subatomic Particle Hideout.}
        \label{em_scaled_spec}
    \end{subfigure}
    \caption{Electromagnetic energy spectrum at the rock-cavern boundary of the Subatomic Particle Hideout: 
    (a) Unweighted
    (b) Energy-weighted.}
    \label{em_energy_spectra}
\end{figure}

The most dominant background from muon-induced secondaries in the rock is the electromagnetic component, consisting of $\gamma$-rays, electrons, and positrons. Their fluxes are listed in Table \ref{em_flux}. The total muon-induced $\gamma$-ray flux will be small compared to the natural $\gamma$-ray background from radioisotopes~\cite{frejus_gamma_ray}, but from Figure \ref{em_energy_spectra} we can see that the energies of muon-induced gammas reach much higher than the typical $\approx$ \SI{3}{MeV} energies from radioisotope decay~\cite{MARTIN20071583}. Additionally, we observe a peak around \SI{0.5}{MeV}, which can be attributed to the annihilation of the positrons in the rock, leading to a single annihilation gamma of \SI{511}{keV} entering the lab. It is also interesting to point out that the neutron capture on $^{23}$Na can lead to a $^{24}$Na nucleus in the first excited state which emits a \SI{472}{keV} $\gamma$-ray. However, inspecting the $\gamma$-ray creation processes extracted from \textsc{geant} shows annihilation is over 800$\times$ stronger in this energy region.

\vspace{0.25cm}
\begin{table}[hb]
\footnotesize
\centering
\renewcommand{\arraystretch}{1.4}
\setlength{\tabcolsep}{5.5pt}
\caption{Simulated muon-induced fluxes for the electromagnetic component.}

\begin{tabular}{l|c|c}
\textbf{Secondary} & \textbf{SPH} [\si{\per\square\metre\per\second}] & \textbf{Cryolab I} [\si{\per\square\metre\per\second}]\\ \hline
Gamma (all)  & 
(5.54 $\pm$ 0.70$_{sys.}$)$\times10^{-1}$ & 
(6.51 $\pm$ 1.06$_{sys.}$) $\times10^{-1}$ \\
 $>$ 0.511 MeV  & 
(2.60 $\pm$ 0.33$_{sys.}$)$\times10^{-1}$ & 
(3.12 $\pm$ 0.51$_{sys.}$) $\times10^{-1}$ \\
 $>$ 2 MeV  & 
(1.59 $\pm$ 0.20$_{sys.}$)$\times10^{-1}$ & 
(1.91 $\pm$ 0.31$_{sys.}$) $\times10^{-1}$ \\
 $>$ 5 MeV  & 
(1.02 $\pm$ 0.13$_{sys.}$)$\times10^{-1}$ & 
(1.22 $\pm$ 0.20$_{sys.}$) $\times10^{-1}$ \\
 $>$ 10 MeV  & 
(6.49 $\pm$ 0.82$_{sys.}$)$\times10^{-2}$ & 
(7.79 $\pm$ 1.27$_{sys.}$) $\times10^{-2}$ \\
Electron & 
(4.39 $\pm$ 0.56$_{sys.}$)$\times10^{-2}$ &
(5.28 $\pm$ 0.86$_{sys.}$)$\times10^{-2}$ \\
Positron & 
(1.08 $\pm$ 0.14$_{sys.}$)$\times10^{-2}$ &
(1.29 $\pm$ 0.21$_{sys.}$)$\times10^{-2}$\\
\end{tabular}

\label{em_flux}
\end{table}

\begin{figure}[htbp]
    \centering
    \begin{subfigure}[b]{1.0\linewidth}
        \centering
        \includegraphics[width=\linewidth]{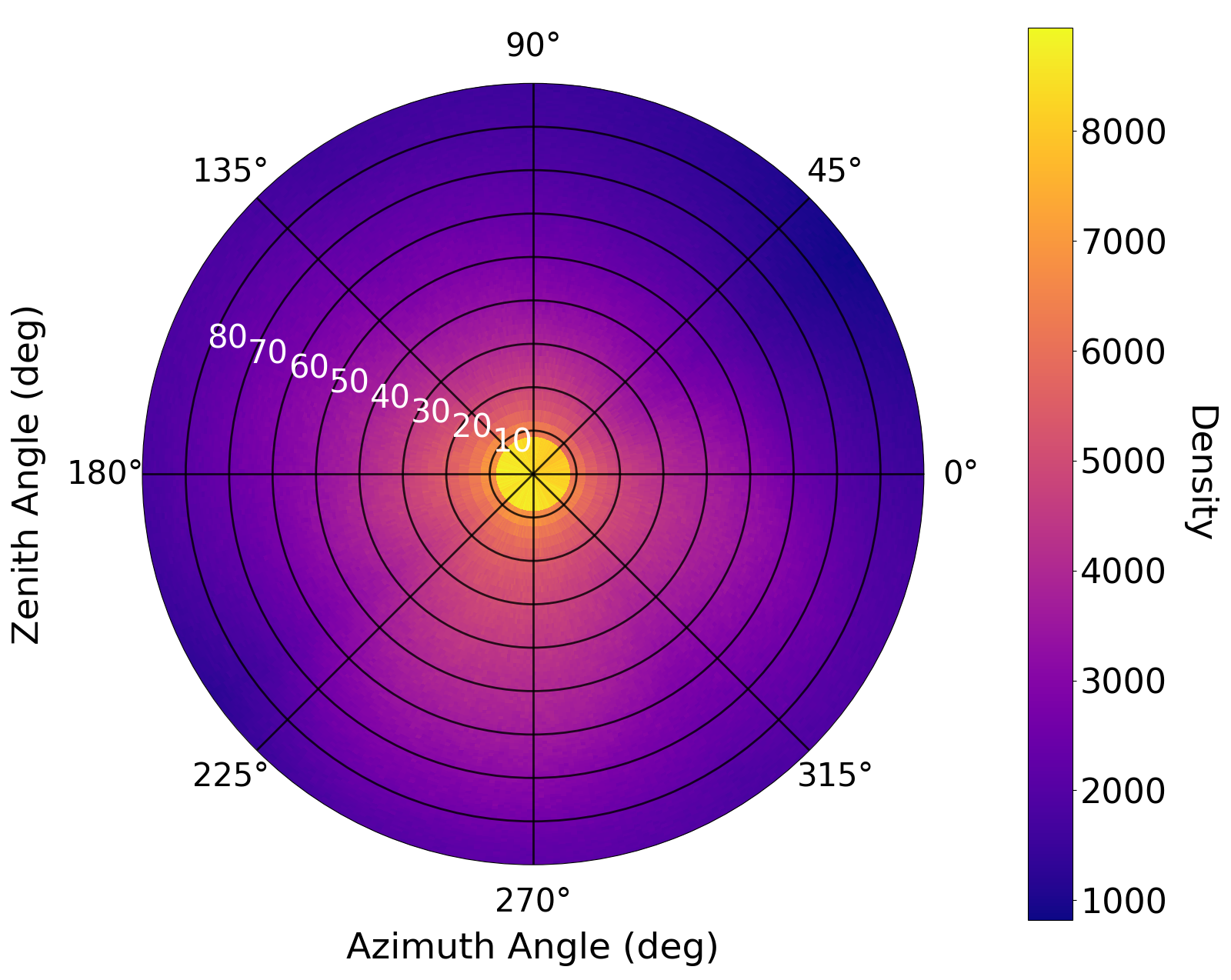}
        \caption{Gamma angular distribution at the rock-cavern boundary for the Subatomic Particle Hideout.}
        \label{hideout_gamma_dist}
    \end{subfigure}

    \vspace{2em}

    \begin{subfigure}[b]{1.0\linewidth}
        \centering
        \includegraphics[width=\linewidth]{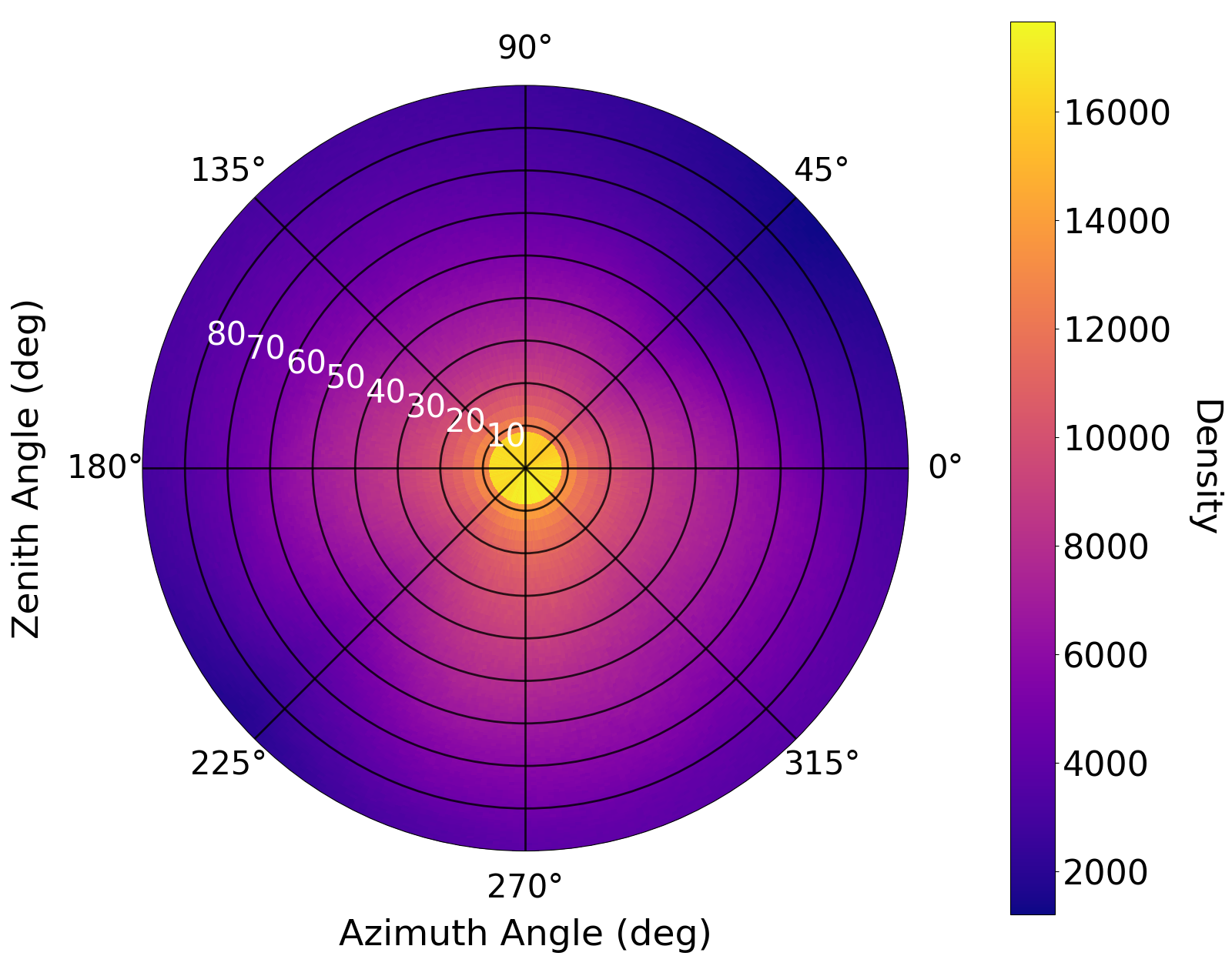}
        \caption{Gamma angular distribution at the rock-cavern boundary for Cryolab I.}
        \label{cryolab_gamma_dist}
    \end{subfigure}

    \caption{Comparison of gamma angular distributions at the rock-cavern boundary: 
    (a) Subatomic Particle Hideout and 
    (b) Cryolab I. Note that \ang{90} corresponds to North in the compass quadrant system.}
    \label{gamma_dist_comparison}
\end{figure}

These rare but energetic secondaries can penetrate shielding and generate additional backgrounds through processes such as pair production, bremsstrahlung, and electromagnetic showers within detector materials. The relative contributions of electrons and positrons are smaller than that of $\gamma$-rays but follow a similar spectral shape, indicating that they largely originate from common muon-induced mechanisms such as photonuclear interactions and muon bremsstrahlung. The angular distributions for the $\gamma$-rays at the rock-cavern boundary can be seen in Figure~\ref{gamma_dist_comparison}, which can information detector shielding and placement needs.

\section{Discussion and Conclusions}\label{sec:Discussion}

For completeness, Table \ref{all_particle_fluxes} lists all relevant secondaries for this study and their fluxes derived from the simulations.

This work presents the first comprehensive Monte Carlo simulation of muon-induced backgrounds for the Colorado Underground Research Institute, establishing critical benchmarks for future low-background experiments at this shallow-underground facility. The simulated muon-induced neutron fluxes are $(8.52 \pm 1.30_{\mathrm{sys}})\times 10^{-3}$ [\si{\per\square\metre\per\second}] and $(8.86 \pm 1.62_{\mathrm{sys}})\times 10^{-3}$ [\si{\per\square\metre\per\second}] for the Subatomic Particle Hideout and Cryolab I, respectively, and are consistent within experimental results from similar depth facilities. The resulting neutron energy spectra exhibit the expected thermal ($\approx \SI{2.5e-8}{MeV}$), epithermal/fast, evaporation (\SI{0.1}{MeV}–\SI{10}{MeV}), and spallation ($\approx \SI{100}{MeV}$) components, with the high-energy tail extending to GeV energies as predicted by muon spallation theory \cite{KUDRYAVTSEV2003688}.

Furthermore, the depth-intensity relation derived in this work extends the Mei \& Hime parameterization to reliably describe muon-induced neutron production in the shallow-underground regime. This predictive capability is essential for the design and characterization of new research facilities operating between the surface and deep-underground environments, where existing models either rely on large extrapolations or fail to capture the rapid increase in neutron flux at small overburdens. By providing a unified parameterization that performs well across the full range of depths, the new relation enables more accurate background estimates and supports the development of intermediate-depth laboratories.

The electromagnetic component dominates the total muon-induced flux, with $\gamma$-ray rates of $(5.54 \pm 0.70_{\mathrm{sys}})\times 10^{-1}$ [\si{\per\square\metre\per\second}] and $(6.51 \pm 1.06_{\mathrm{sys}})\times 10^{-1}$ [\si{\per\square\metre\per\second}] for the two research spaces, exceeding neutron fluxes by nearly two orders of magnitude and extending into \SI{}{GeV} energies. 
The ability of these high-energy photons to penetrate conventional lead shielding and induce electromagnetic cascades within detector materials necessitates careful consideration in experimental design, particularly for rare-event searches sensitive to backgrounds in the MeV energy range. Our results provide essential input parameters for Monte Carlo modeling of detector responses and optimization of active and passive veto systems for future CURIE experiments.

The demonstrated 7.6\% discrepancy between mean energy approximations and angular-dependent muon energy sampling highlights a critical methodological consideration for underground background simulations. Previous studies have noted similar systematic effects \cite{REICHHART201367, Empl_2014}, but our work provides the first quantitative assessment at CURIE's specific depth and topological setting. This finding has particular relevance for facilities with complex overburden profiles, where azimuthal anisotropies in the muon flux cannot be neglected. The site-specific corrections for laboratory geometry and the transition from point-source to finite-volume treatment represent important refinements that improve simulation fidelity without significant computational overhead.

The successful integration of the \textsc{mute} and \textsc{geant4} frameworks demonstrates a robust methodology for end-to-end underground background characterization that can be readily adapted and provides a computationally efficient approach that could be adopted by other underground facilities. Therefore, we have made this computational framework available online at 
\href{https://doi.org/10.5281/zenodo.17196581}{https://doi.org/10.5281/zenodo.17196581}~\cite{zenodo}, in order for future work to extend this analysis to other underground facilities to establish standardized simulation protocols and enable meaningful inter-facility comparisons for the broader low-background physics community.

\begin{table*}[ht]
\footnotesize
\centering
\renewcommand{\arraystretch}{1.7}
\setlength{\tabcolsep}{10.0pt}
\caption{Primary muon and the simulated muon-induced fluxes for all secondaries at Subatomic Particle Hideout and Cryolab I. The primary muon flux presented for SPH and Cryolab I are taken from \cite{dakota}.}

\begin{tabular}{l|c|c}
 & \textbf{SPH} [\si{\per\square\metre\per\second}] & \textbf{Cryolab I} [\si{\per\square\metre\per\second}]\\ \hline
Primary Muon & (2.39 $\pm$ 0.25$_{sys.}$)$\times10^{-1}$ & (2.59 $\pm$ 0.26$_{sys.}$)$\times10^{-1}$ \\
\textbf{Secondary} &  &   \\
$\gamma$ & (5.54 $\pm$ 0.70$_{sys.}$)$\times10^{-1}$ & (6.51 $\pm$ 1.06$_{sys.}$)$\times10^{-1}$ \\
$e^{-}$ & (4.39 $\pm$ 0.56$_{sys.}$)$\times10^{-2}$ & (5.28 $\pm$ 0.86$_{sys.}$)$\times10^{-2}$ \\
$e^{+}$ & (1.08 $\pm$ 0.14$_{sys.}$)$\times10^{-2}$ & (1.29 $\pm$ 0.21$_{sys.}$)$\times10^{-2}$ \\
$\bar{\nu}_e$ & (6.37 $\pm$ 1.04$_{sys.}$)$\times10^{-3}$ & (8.98 $\pm$ 1.47$_{sys.}$)$\times10^{-3}$ \\
$n$ & (8.52 $\pm$ 1.30$_{sys.}$)$\times10^{-3}$ & (8.86 $\pm$ 1.62$_{sys.}$)$\times10^{-3}$ \\
$\nu_e$ & (3.39 $\pm$ 0.55$_{sys.}$)$\times10^{-3}$ & (5.78 $\pm$ 0.94$_{sys.}$)$\times10^{-3}$ \\
$\nu_\mu$ & (8.47 $\pm$ 1.39$_{sys.}$)$\times10^{-4}$ & (1.39 $\pm$ 0.23$_{sys.}$)$\times10^{-3}$ \\
$\bar{\nu}_\mu$ & (8.46 $\pm$ 1.38$_{sys.}$)$\times10^{-4}$ & (1.39 $\pm$ 0.23$_{sys.}$)$\times10^{-3}$ \\
$p$ & (1.14 $\pm$ 0.19$_{sys.}$)$\times10^{-4}$ & (1.32 $\pm$ 0.22$_{sys.}$)$\times10^{-4}$ \\
$\pi^{+}$ & (8.57 $\pm$ 1.40$_{sys.}$)$\times10^{-5}$ & (9.65 $\pm$ 1.58$_{sys.}$)$\times10^{-5}$ \\
$\pi^{-}$ & (8.38 $\pm$ 1.37$_{sys.}$)$\times10^{-5}$ & (9.53 $\pm$ 1.56$_{sys.}$)$\times10^{-5}$ \\
\end{tabular}

\label{all_particle_fluxes}
\end{table*}

\section*{Acknowledgments}\label{sec:Acknowledgments}
This material is based upon work supported by a National Science Foundation Graduate Research Fellowship under Grant No. DGE-2137099, a U.S. Department of Energy Office of Science Grant No. DE-FG02-93ER40789, and the Colorado School of Mines via faculty start up funds and the ARCS Foundation. We would like to thank the Colorado School of Mines Geology Department and Filip Kasprowicz for their work on the rock sample analysis.

\clearpage
\bibliography{main.bib}

\end{document}